\documentclass{aa}  
\def\ha{H$\alpha$}
\def\hb{H$\beta$}
\def\nii{[N\,{\sc ii}]}
\def\oiii{[O\,{\sc iii}]}
\def\cab{{CaII H\&K}}
\def\car{{CaII IR trip.}}
\def\cabr{{CaII H\&K + IR trip.}}
\def\rha{$r$$-$H$\alpha$}
\def\ri{$r$$-$$i$}
\def\jh{$J$$-$$H$}

\def\magcir{\ \raise-2.truept\hbox{\rlap{\hbox{$\sim$}}\raise5.truept
    \hbox{$>$}\ }}
\def\kms{\relax \ifmmode {\,\rm km\,s}^{-1}\else \,km\,s$^{-1}$\fi}
\usepackage{graphicx}
\usepackage{longtable,lscape}
\usepackage{txfonts}
\begin{document}
%
\title{IPHAS and the symbiotic stars}\subtitle{II. New discoveries and
  a sample of the most common mimics \thanks{Based on observations
    obtained at; the 2.6m Nordic Optical Telescope operated by NOTSA;
    the 2.5m~INT and 4.2m~WHT telescopes of the Isaac Newton Group of
    Telescopes in the Spanish Observatorio del Roque de Los Muchachos
    of the Instituto de Astrof\'\i sica de Canarias; the 2.3m~ANU 
    telescope at Siding Spring Observatory, Australia; the Asiago
    1.82m telescope of the INAF Astronomical Observatory of Padova,
    Italy; and the 2.1m telescope at San Pedro Martir, Mexico. This
    publication makes use of data products from the Two Micron All Sky
    Survey, which is a joint project of the University of
    Massachusetts and the Infrared Processing and Analysis
    Center/California Institute of Technology, funded by the National
    Aeronautics and Space Administration and the National Science
    Foundation. This research has also made use of the SIMBAD
    database, operated at CDS, Strasbourg, France.}}

\author{R.L.M. Corradi\inst{1,2}
          \and
        M. Valentini\inst{3,4}
           \and
        U. Munari\inst{3}
          \and
	J.E. Drew\inst{5}
          \and
        E.R. Rodr\'\i guez--Flores\inst{6,1}
          \and
        K. Viironen\inst{1,2}
         \and
        R. Greimel\inst{7}
         \and
        M. Santander--Garc{\'{\i}}a\inst{8,1,2}
	\and
        L. Sabin\inst{9}
          \and
        A. Mampaso\inst{1,2}
         \and
        Q. Parker\inst{10,11}
         \and
        K. De Pew\inst{10}
         \and
        S.E. Sale\inst{12}
        \and
        Y.C. Unruh\inst{12}
        \and
        J.S. Vink\inst{13}
        \and
        P. Rodr{\'{\i}}guez--Gil\inst{8,1,2}
        \and
        M.J. Barlow\inst{14}
        \and
        D.J. Lennon\inst{15}
        \and
        P.J. Groot\inst{16}
        \and
        C. Giammanco\inst{1,2}
        \and
        A.A. Zijlstra\inst{17}
        \and
        N.A. Walton\inst{18}
	}

   \offprints{R. Corradi}

   \institute{
Instituto de Astrof{\'{\i}}sica de Canarias, E-38200 La Laguna, 
Tenerife, Spain \email{rcorradi@iac.es}
   \and
Departamento de Astrof{\'{\i}}sica, Universidad de La Laguna, 
E-38205 La Laguna, Tenerife, Spain 
\and 
INAF, Osservatorio Astronomico di Padova, via dell'Osservatorio 8, 
36012 Asiago (VI), Italy
\and
Department of Astronomy, University of Padova, Asiago Astrophysical
Observatory, 36012 Asiago (VI), Italy
   \and
Centre for Astrophysics Research, STRI, University of Hertfordshire,
College Lane, Hatfield, AL10 9AB, UK
   \and
Instituto de Geof\'\i sica y Astronom\'\i a, Calle 212, N. 2906, 
CP 11600, La Habana, Cuba
   \and
Institut f\"ur Physik, Karl-Franzen Universit\"at Graz, 
Universit\"atsplatz 5, 8010 Graz, Austria
   \and
Isaac Newton Group of Telescopes, Apart. de Correos 321, 
38700 Santa Cruz de la Palma, Spain
   \and
Jodrell Bank Centre for Astrophysics, Alan Turing Building, 
University of Manchester, Manchester, M13 9PL, UK
   \and
Department of Physics, Macquarie University, Sydney, NSW 2109, Australia
   \and
Anglo-Australian Observatory, PO Box 296, Epping, NSW 1710, Australia
   \and
Astrophysics Group, Imperial College of Science, Blackett Laboratory, Prince 
Consort Road, London SW7 2AZ, UK
   \and
Armagh Observatory, College Hill, Armagh BT61 9DG, Northern Ireland
   \and
Department of Physics and Astronomy, University College London, Gower Street, 
London WC1E 6BT, UK
   \and
Space Telescope Science Institute, 3700 San Martin Drive, Baltimore, 
MD 21218, USA
   \and
Department of Astrophysics, IMAPP, Radboud University Nijmegen, PO Box 9010, 
6500 Gl Nijmegen, the Netherlands
   \and
Jodrell Bank Centre for Astrophysics, School of Physics and Astronomy,
University of Manchester, Oxford Street, Manchester M13 9PL, UK
\and
Institute of Astronomy, Cambridge University, Madingley Road, Cambridge, 
CB3 0HA, UK
             }

\date{Received ... / Accepted ...}

\abstract{Knowledge of the total population of symbiotic stars in
  the Galaxy is important for understanding basic aspects of stellar
  evolution in interacting binaries and the relevance of this class of
  objects in the formation of supernovae of type Ia.}
{In a previous paper, we presented the selection
  criteria needed to search for symbiotic stars in IPHAS, the INT \ha\ survey
  of the Northern Galactic plane. IPHAS gives us the opportunity to
  make a systematic, complete search for symbiotic stars in a
  magnitude-limited volume.}  
{Follow-up spectroscopy at different telescopes worldwide of a sample
  of sixty two symbiotic star candidates is presented.}
{Seven out of nineteen S-type candidates observed spectroscopically are 
confirmed to be genuine symbiotic stars.  The spectral type of their red giant
components, as well as reddening and distance, were computed by modelling 
the spectra.  Only one new D-type symbiotic system, 
out of forty-three candidates observed, was found. 
This was as expected (see discussion in our paper 
  on the selection criteria). The
  object shows evidence for a high density outflow expanding at a
  speed $\ge$65~\kms.

Most of the other candidates are lightly reddened classical T Tauri
stars and more highly reddened young stellar objects that may be
either more massive young stars of HAeBe type or classical Be
stars. In addition, a few notable objects have been found, such as three
new Wolf-Rayet stars and two relatively high-luminosity evolved massive
stars.
We also found a helium-rich source, possibly a dense ejecta hiding a WR star, 
which is surrounded by a large ionized nebula.}  
{These spectroscopic data allow us
  to refine the selection criteria for symbiotic stars in the IPHAS
  survey and, more generally, to better understand the behaviour of
  different \ha\ emitters in the IPHAS and 2MASS colour-colour
  diagrams.}  
\keywords{ Surveys; (Stars:) binaries: symbiotic; Stars:
  emission-line, Be; Stars: pre-main sequence; Stars: Wolf-Rayet;
  (ISM:) planetary nebulae: general}

\titlerunning{IPHAS and the symbiotic stars. II.}
\authorrunning{R.L.M. Corradi et al.}
\maketitle


\begin{table*}[!ht]
\caption{The new symbiotic stars and their parameters as determined in
  this paper.}
\begin{tabular}{lcccrrrccccccc}
\hline\hline
Name (IPHASJ....)  & r$^\ast$      & i$^\ast$   & \ha$^\ast$    &\multicolumn{1}{c}{J} & \multicolumn{1}{c}{H} & \multicolumn{1}{c}{K}  & IR   & Spec. & E(B-V) &
A(V) & A(K) & M(K) & d\\
                   &  [mag] &     &        &        &      &                                                       & type & type  &        &
     &      &      & [kpc]\\
\hline\\[-7pt]                                                                   
182906.08--003457.2& 17.53 & 14.87 & 15.66 & 10.97 & 9.61 & 9.05 & S  & M5.7 III & 1.30 &  5.08 &  0.59 &$-$5.65 & 6.7 \\      
183501.83+014656.0 & 16.34 & 14.47 & 13.83 & 10.74 & 9.48 & 8.92 & S  & M5.5 III & 1.40 &  5.45 &  0.63 &$-$5.55 & 5.9 \\   
184446.08+060703.5 & 14.74 & 13.34 & 12.70 & 11.04 & 9.88 & 9.40 & S  & M2.0 III & 1.00 &  3.76 &  0.44 &$-$3.80 & 3.6 \\   
184733.03+032554.3 & 18.55 & 15.32 & 16.13 & 10.07 & 8.44 & 7.71 & S  & M6.5 III & 2.20 &  8.71 &  0.96 &$-$6.22 & 3.9 \\   
185323.58+084955.1 & 16.53 & 14.25 & 14.28 & 10.32 & 9.05 & 8.51 & S  & M6.5 III & 1.35 &  5.33 &  0.61 &$-$6.15 & 6.5 \\   
193436.06+163128.9 & 16.84 & 14.65 & 14.49 & 10.70 & 9.34 & 8.77 & S  & M5.9 III & 1.08 &  4.23 &  0.49 &$-$5.75 & 6.4 \\   
193501.31+135427.5 & 14.04 & 12.51 & 12.81 & 10.05 & 8.98 & 8.58 & S  & M1.5 III & 0.60 &  2.25 &  0.27 &$-$3.66 & 2.5 \\   
194607.52+223112.3 & 17.54 & 16.04 & 14.70 &  9.22 & 7.19 & 5.96 & D  &          &      &       &       &        &     \\
\hline                                                                                                              
\end{tabular}
\vspace*{0.2cm}
\newline
$^\ast$ From \cite{w08}.
\label{T-symbio}
\end{table*}

\section{Introduction}

Symbiotic stars, the interacting binaries with the longest periods,
are recognised as key objects in the study of different aspects of stellar
evolution. These include the formation of supernovae of type Ia
(Munari \& Renzini 1992, Hachisu et al. 1999), the powering mechanism
of supersoft X-ray sources (cf. Jordan et al. 1996) and the formation of
jets (Tomov 2003) and of bipolar (planetary) nebulae (\cite{c03}).

The testing of these possible roles, and in particular their ability
to form SNe Ia, depends on the total population of symbiotic stars within
galaxies. This figure is, however, poorly known, even in our own Galaxy
where less than 200 systems are known (\cite{b00}), out of a predicted
total population that ranges from 3$\times$$10^3$ (\cite{a84}) up to
4$\times$$10^5$ (\cite{mcm03}).

To improve the situation, a comprehensive search for
symbiotic stars in the part of the Galactic plane visible from the
northern hemisphere was started (Corradi et al. 2008, hereafter paper
I).  The search begins with analysing the data from the INT Photometric
\ha\ Survey of the northern Galactic plane (IPHAS, \cite{d05}), taking
advantage of the generally strong \ha\ emission and presence of a cool
giant that characterises this class of objects.  Combination of these
properties with constraints from 2MASS near--IR data defines the
selection criteria adopted in paper I. There, a list of about one
thousand symbiotic star candidates was drawn up from the set of IPHAS
\ha\ emitters available at that time (Witham et al. 2008). As shown 
in paper I, this list suffers from significant contamination by 
reddened Be stars and young stellar objects, which are frequent in the 
Galactic plane.

In this paper, we present follow-up spectroscopy for a sample of
62 IPHAS candidate symbiotic stars. Observations are presented in
Sect.~2. Section~3 is a discussion of the spectroscopically confirmed 
new symbiotic systems, while Sect.~4 outlines the majority 
group of mimics revealed in our spectra: among these there are several 
notable, rare object types. Discussion and perspectives are presented in 
Sect.~5.


\begin{figure*}[!ht]
\centering
\includegraphics[width=17cm]{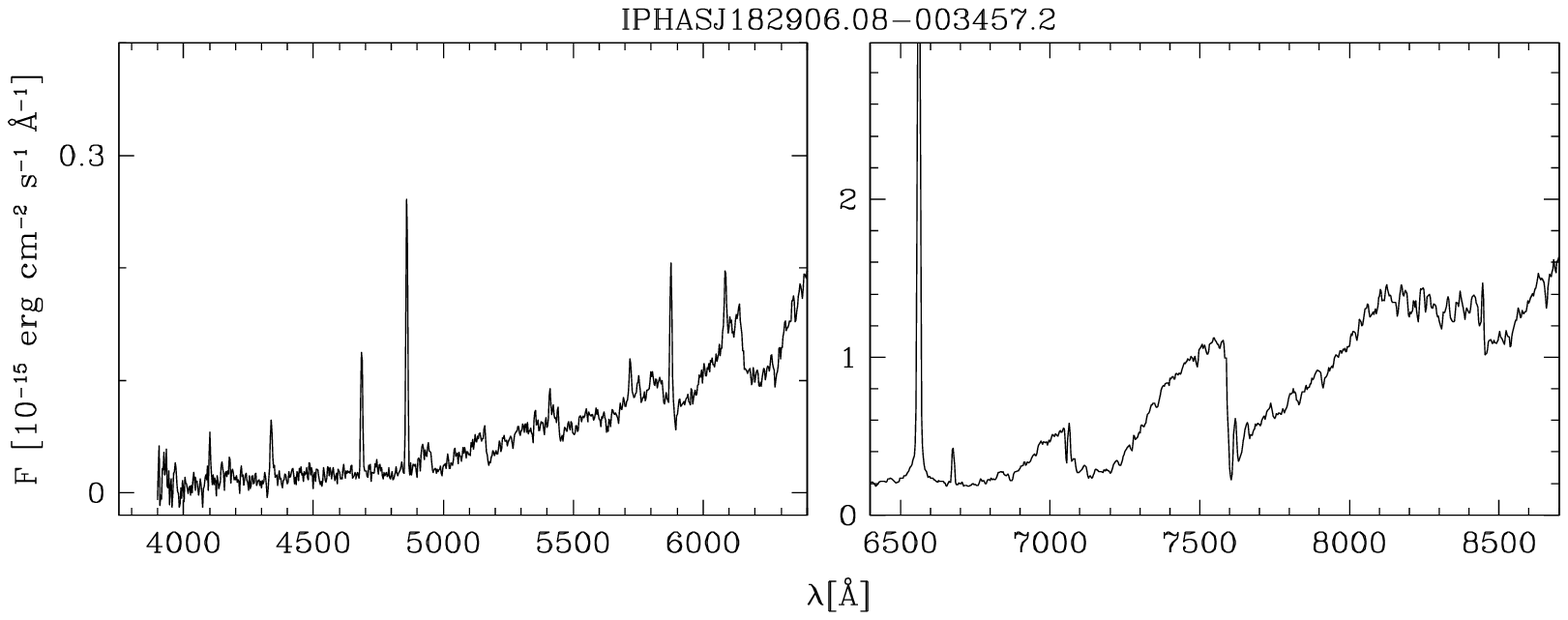}\\[5pt]
\includegraphics[width=17cm]{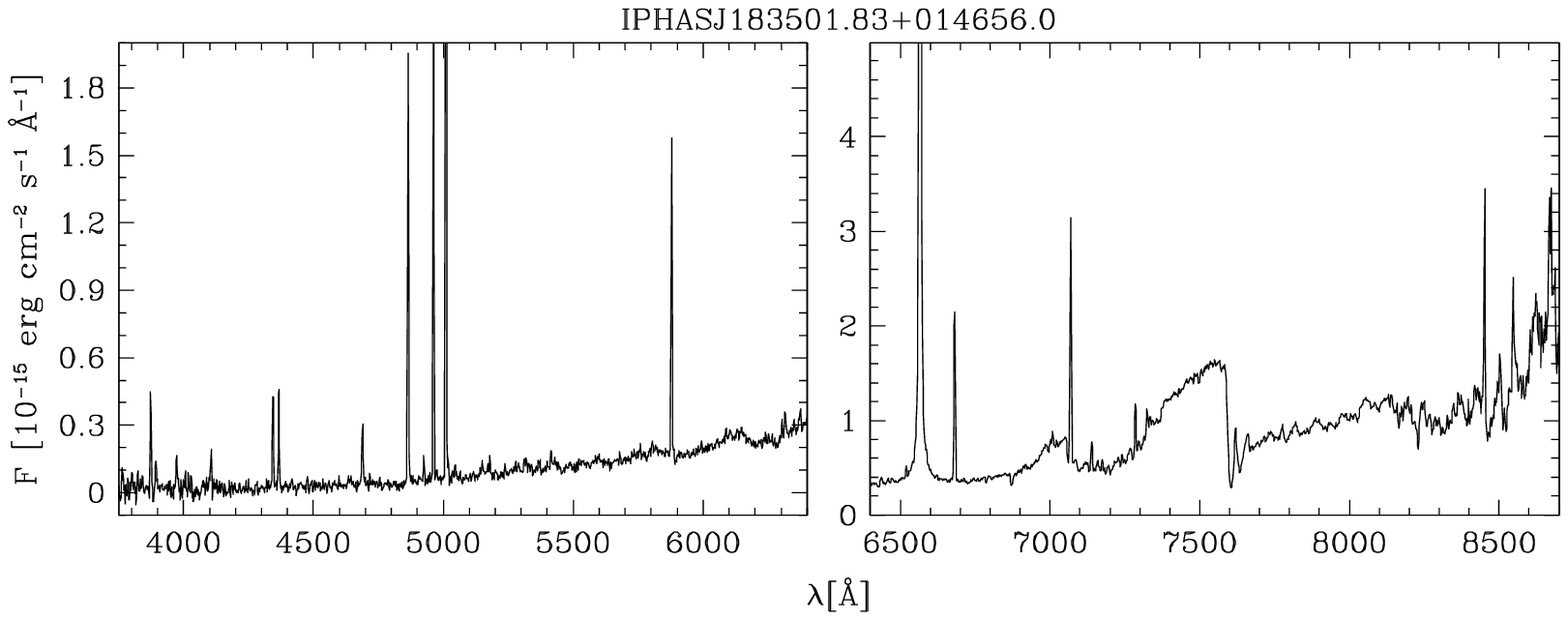}\\[5pt]
\includegraphics[width=17cm]{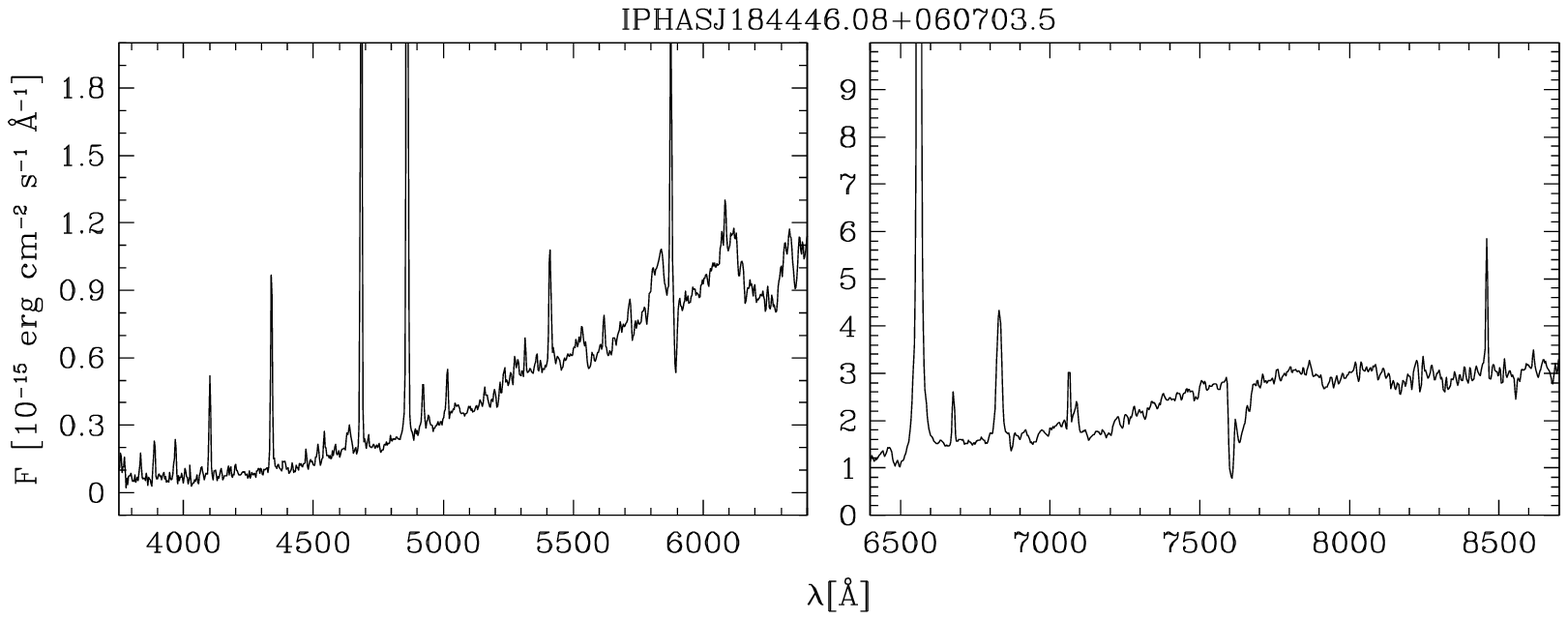}
\caption{Spectra of the new S-type symbiotic stars
  IPHASJ182906.08--003457.2 (top), IPHASJ183501.83+014656.0 (middle),
  and IPHASJ184446.08+060703.5 (bottom). The blue and red regions of
  the spectra are displayed with different intensity cuts to show both
  faint and bright features.}
\label{F-specsymbio1}
\end{figure*}

\begin{figure*}[!ht]
\centering
\includegraphics[width=17cm]{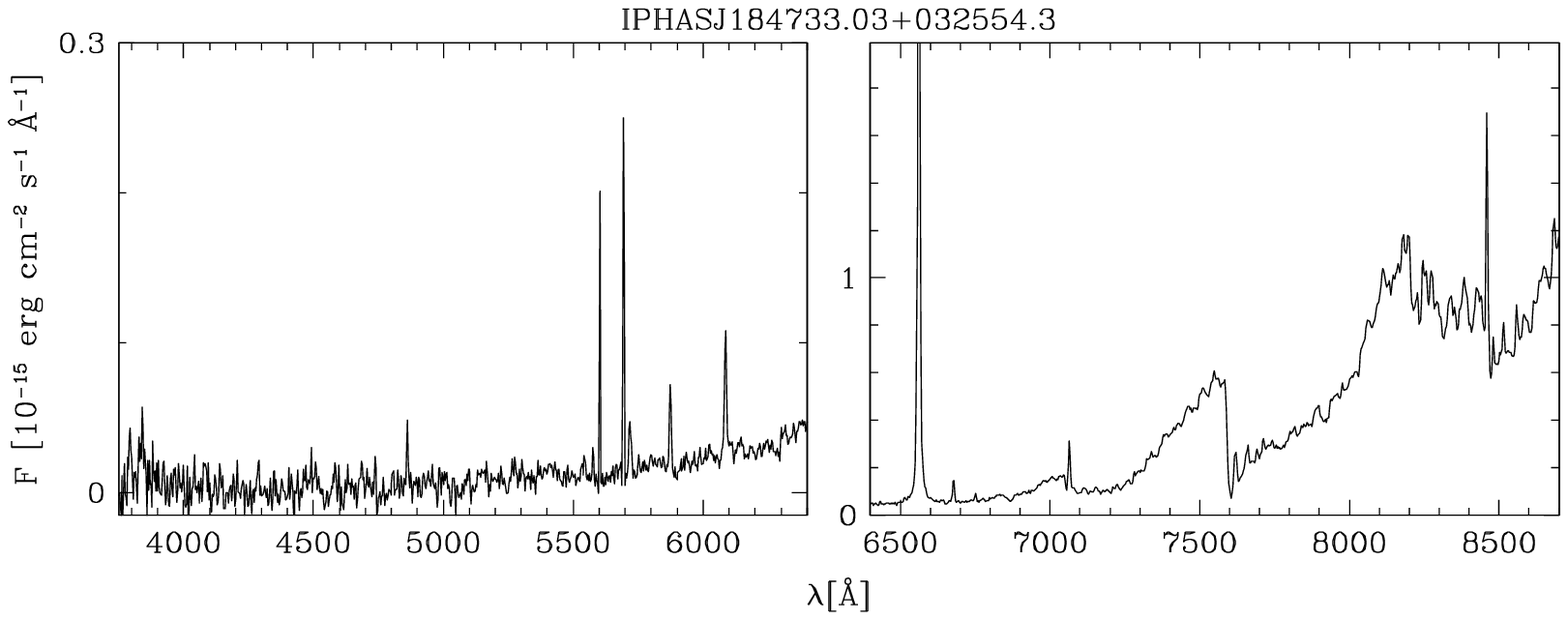}\\[5pt]
\includegraphics[width=17cm]{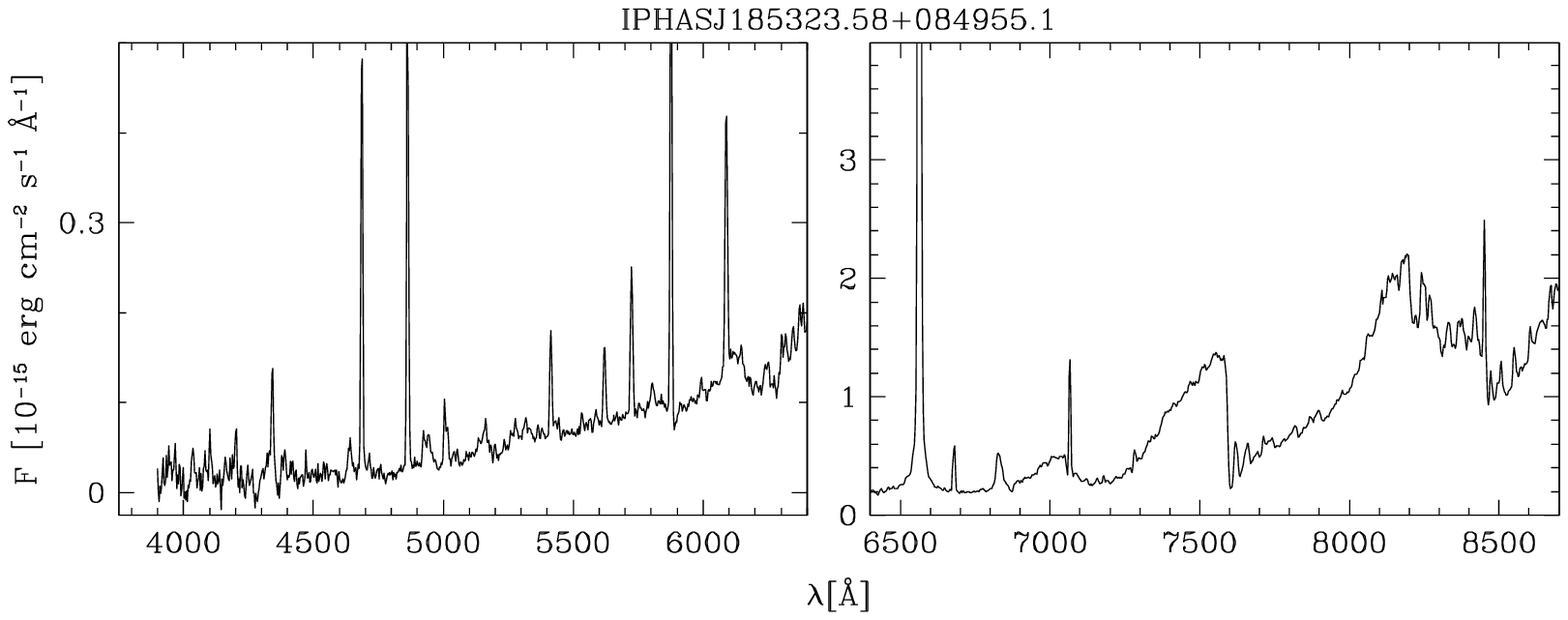}\\[5pt]
\includegraphics[width=17cm]{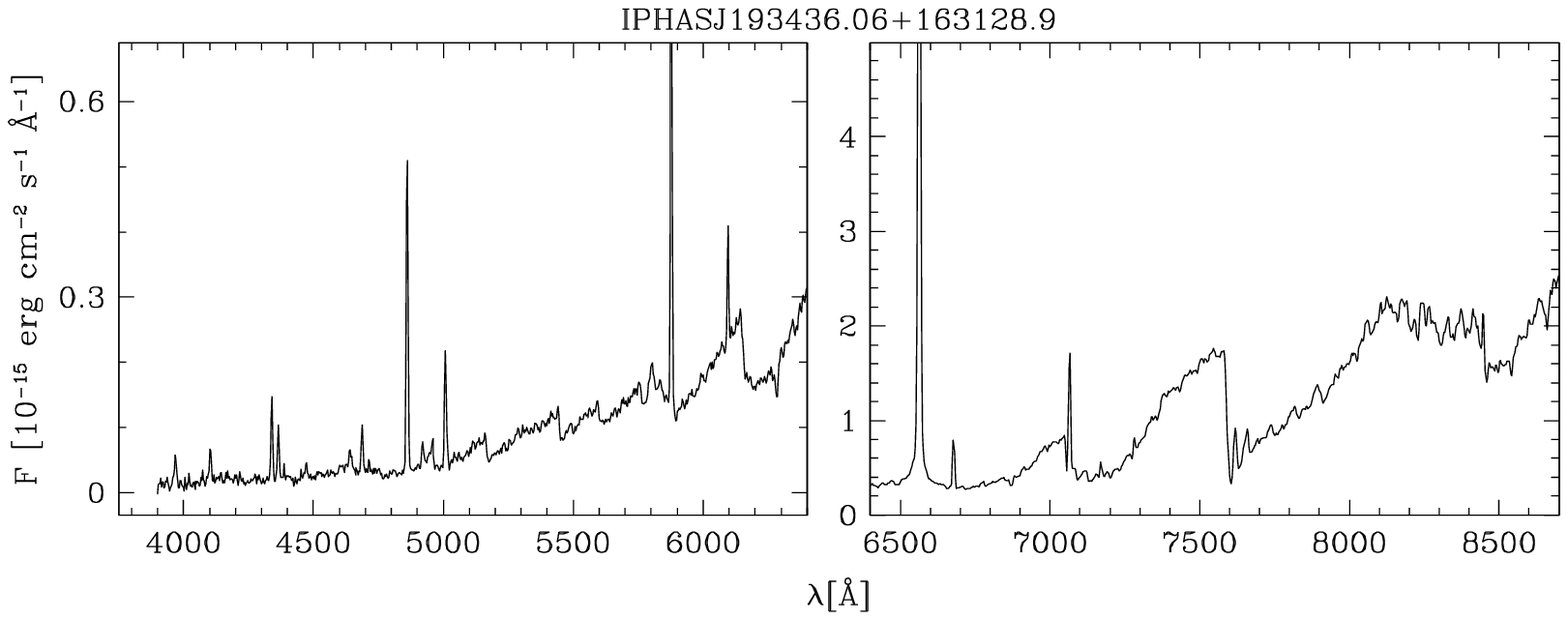}
\caption{Spectra of the new S-type symbiotic stars
 IPHASJ184733.03+032554.3 (top), IPHASJ185323.58+084955.1 (middle), and
 IPHASJ193436.06+163128.9 (bottom).}
\label{F-specsymbio2}
\end{figure*}

\begin{figure*}[!ht]
\centering
\includegraphics[width=17cm]{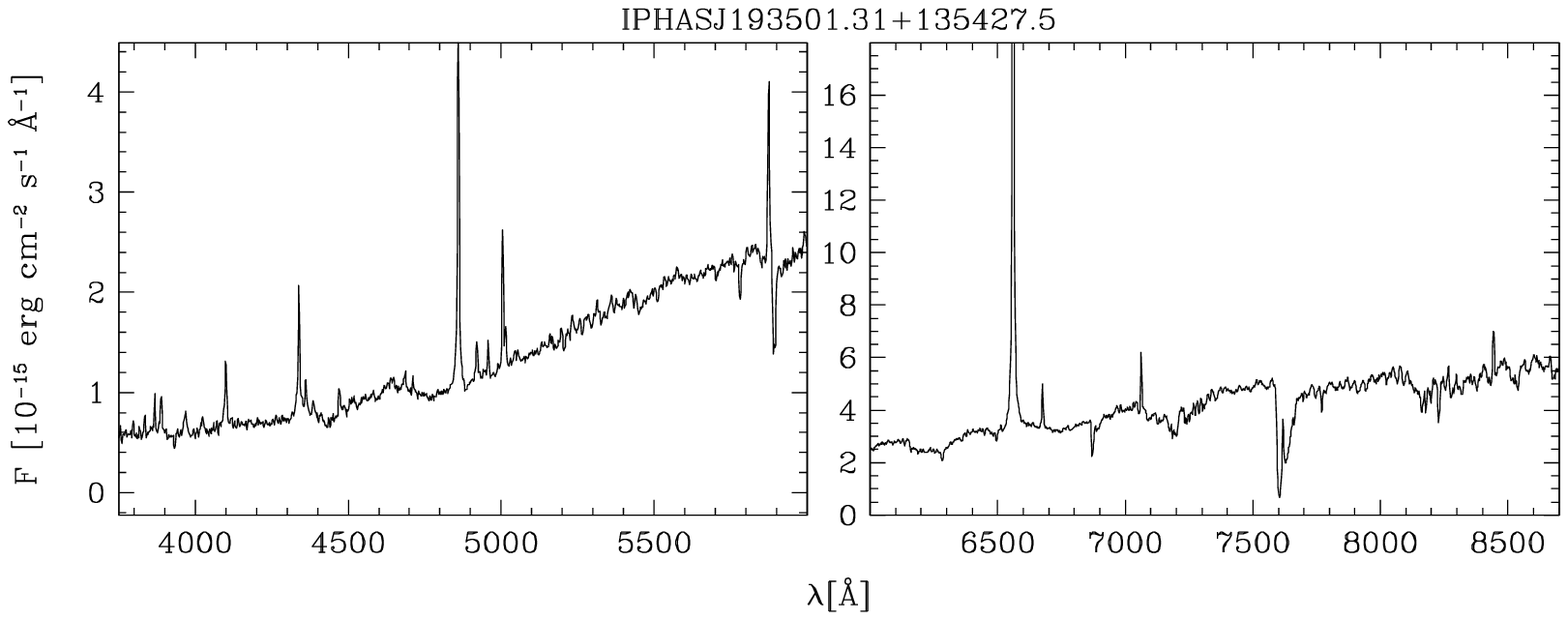}\\[5pt]
\includegraphics[width=17cm]{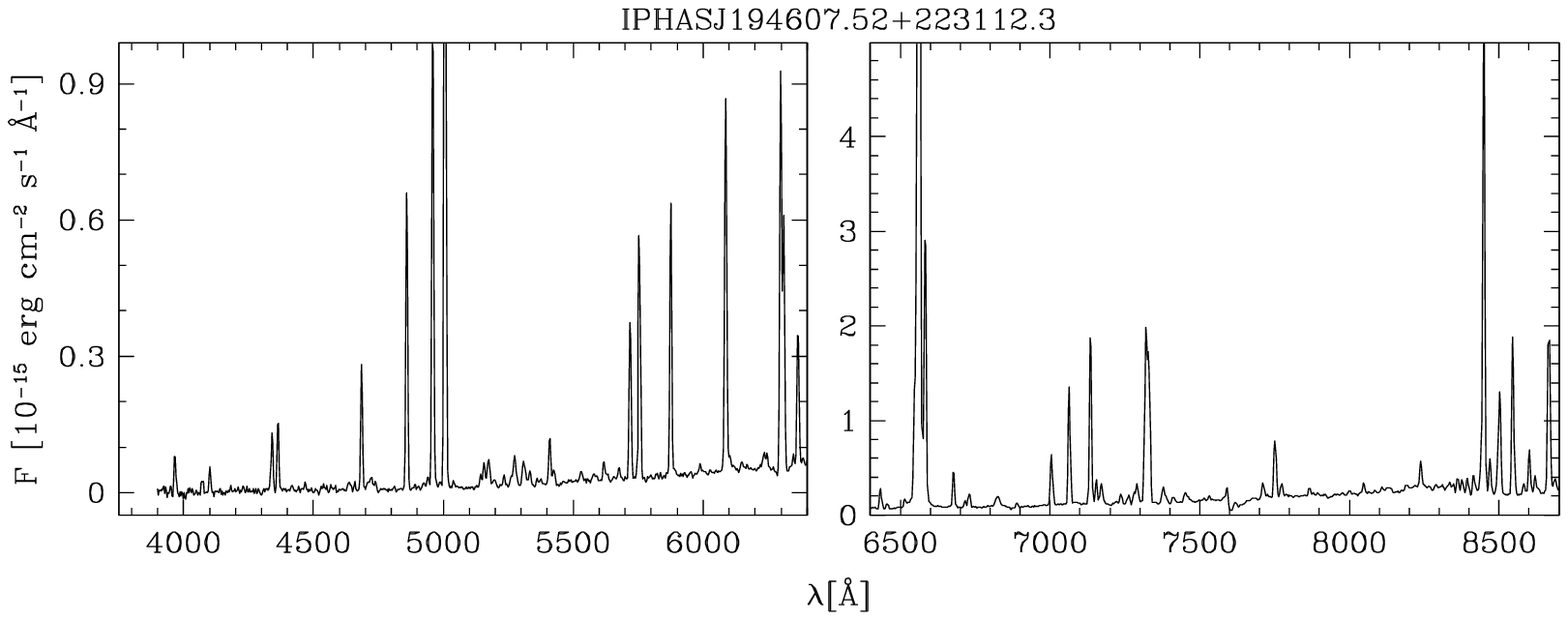}
\caption{Spectra of the S-type new symbiotic star
  IPHASJ193501.31+135427.5 (top), and of the D-type new symbiotic star
  IPHASJ194607.52+223112.3 (bottom).}
\label{F-specsymbio3}
\end{figure*}

\section{Spectroscopic observations}

In paper I, the discovery of three new symbiotic stars from
exploratory spectroscopic observations obtained at the 2.5m Isaac
Newton Telescope (INT) on La Palma, Spain, was presented.  Since then,
our spectroscopic campaign has progressed on different telescopes
worldwide in various runs from June 2006 to June 2008.

Six out of the eight newly discovered symbiotic stars presented in the next
section were observed at the 2.6m Nordic Optical Telescope on La Palma in 
September 2007. The ALFOSC spectrograph was used, in combination with grism
\#4 and a 0$''$.5 wide slit.  This set-up provides a reciprocal 
dispersion of 3.1~\AA\ per pixel, a resolution of 8.1~\AA, and a spectral 
coverage between 3700 and 8700~\AA.
The 2kx2kEEV ALFOSC CCD\#8 suffers from severe fringing in the 
red, with an amplitude as large as 25\%\ at 8000~\AA. This is only partially 
corrected using lamp flat fields taken with the telescope at the position of 
the target. Exposure times ranged between 10 and 30 min per object. One 5~min
higher resolution spectrum of IPHASJ194607.52+223112.3 was also obtained at
the NOT using grism\#17, covering the region 6350-6850~\AA\ with a 0.65~\AA\
resolution.

Other spectra were obtained at the INT using the IDS
spectrograph. They include the new symbiotic star
IPHASJ183501.83+014656.0 observed in September 2006.  Grating R300V
was used, which gives a reciprocal dispersion of 1.9~\AA\ per pixel,
and a spectral coverage from 3700 to 8700~\AA\ (these figures slightly
vary from night to night).  The slit width was 1$''$.1, providing a
spectral resolution of 5.0~\AA. Typical exposure times were 20 to 30
min per object. The 2kx4k IDS EEV CCD also suffers from 
fringing redward of $\sim$7000~\AA. In addition, flux calibration is
uncertain above 8000~\AA\ because of significant optical aberrations
at the edge of the large format CCD used with IDS.

The new symbiotic star IPHASJ193501.31+135427.5, together with another
twenty candidates, was observed at the 2.3m ANU telescope
at Siding Spring Observatory, Australia, with the DBSB instrument, in
August 2008. Blue spectra covering the region from 3700 to 6000~\AA,
at a reciprocal dispersion of 2.0~\AA\ per pixel and a resolution of
4.5~\AA\ were obtained with exposure times of 30 min.  Red spectra
covering the range 5700-9000~\AA\ were also obtained, with the same
exposure time and similar dispersion and resolution (1.9~\AA\ per
pixel and 5.0~\AA, respectively).

The other spectra were obtained at the 4.2m WHT telescope on La Palma with
the ISIS spectrograph; the 2.1m telescope at San Pedro Martir, Mexico with
the Boller \& Chivens spectrograph; and the Asiago 1.8m telescope of
the INAF Padova Astronomical Observatory equipped with the AFOSC
imager+spectrograph. Resolution, depth and spectral coverage varied
depending on the instruments and optical elements used in the different
nights. In all cases, the reciprocal dispersion was between 1.6~\AA\ and
4.0~\AA\ per pixel, and spectral resolution between 4 and 9~\AA. The
spectral coverage is indicated in the tables in the following sections of
the paper.


Several spectrophotometric standards were observed during the nights for
relative flux calibration. Reduction of all spectra was performed with 
IRAF in a standard fashion.

\section{New IPHAS symbiotic stars}

Seven out of nineteen S-type candidate symbiotic stars observed,
and one out of forty-three D-type candidates, turned out to be new
symbiotic stars.
Their names (according to IAU--approved IPHAS nomenclature),
IPHAS\footnote{IPHAS magnitudes are tied to the system of
  corresponding SDSS bands, with zero magnitude defined by the SED of
  Vega.}  and 2MASS magnitudes, near-IR type, and other parameters
obtained as discussed below, are listed in Tab.~\ref{T-symbio}. Their
optical spectra are shown in Figs.~\ref{F-specsymbio1} to
\ref{F-specsymbio3}.  Note that in order to accept a source as a
genuine symbiotic star we opted in favor of the original
classification criterion due to Allen (1984), which requires the
presence of both a cool giant and ionization condition high enough to
produce at least HeII emission lines.  This criterion isolates an
homogeneous set of objects that are characterised by stable H-burning
conditions on the surface of the white dwarf accreting from the cool
giant companion, a scenario theoretically proposed by Tutukov and
Yungelson (1976), and Paczynski and Rudak (1980), and confirmed
observationally by Munari and Buson (1994), and Sokoloski (2003).  The
more relaxed classification criterion favoured by Belczy\'nski et al.
(2000) does not isolate a similarly homogeneous group.

A brief description of the spectrum of each object follows.

{\it IPHASJ182906.08-003457.2} (Fig.~\ref{F-specsymbio1}, top). It
shows a classical spectrum of a symbiotic star, with prominent red
continuum with the deep TiO absorption bands typical of an M
giant\footnote{The heads of the most prominent TiO bands 
in the range covered by our spectra are at 7053, 7589,
8206, and 8432~\AA\ (see \cite{k91}).}, and emission lines
from low to high excitation, including: the HI Balmer series down to
H$\gamma$, HeI 5875, 6678, 7065, HeII 4686, 5411, faint [FeVII]5721,
[CaV]6086 or [FeVII] 6087, and faint OI 8446. No other nebular
forbidden lines are detected at the depth of our spectra. The
\ha/\hb\ flux ratio is as large as 24.

{\it IPHASJ183501.83+014656.0} (Fig.~\ref{F-specsymbio1},
middle). Included in the list of faint \ha\ emitters by Robertson and
Jordan (1989), our spectrum reveals its symbiotic nature by showing
both a rising continuum with TiO bands and a rich emission line
spectrum. This includes [NeIII]3869, the HI Balmer series down to
H$\delta$, [OIII]4363 stronger than H$\delta$, HeII4686, [OIII]5007
three times stronger than \hb\ and 11 times stronger than [OIII]4363,
[NII]5755 stronger than [NII]6583, HeI 5875,6678,7065,7281, and
OI8446. The \ha/\hb\ ratio is also 24.

{\it IPHASJ184446.08+060703.5} (Fig.~\ref{F-specsymbio1}, bottom). The
object was originally included in the catalogue of planetary nebulae
(PNe) by Kohoutek (1965) with name PN K3-12. However, it was suspected
to be a symbiotic star by Acker \& Stenholm (1990). Our spectrum
confirms its symbiotic nature, revealed by the emission line spectrum
combined with strong 6825 and 7082 Raman scattered lines and, albeit
shallow, TiO bands. Emission lines include the HI Balmer series, HeI
5875,6678,7065, HeII 4686 almost as strong as \hb, HeII~5411,
FeII~4923, 5018, faint [FeVII]~5721 and weak [CaV]~6086 or
[FeVII]~6087.

{\it IPHASJ184733.03+032554.3} (Fig.~\ref{F-specsymbio2}, top).  The
object seems very reddened, with no signal below \hb.  Its symbiotic
nature is revealed by the strong TiO bands and the emission lines
which include [FeVII]~5720, HeI~5875, 6678, 7065, [CaV]~6086 or
[FeVII]~6087, OI8446 and \ha\ which is 110 times stronger than \hb.

{\it IPHASJ185323.58+084955.1} (Fig.~\ref{F-specsymbio2},
middle). Another classical spectrum of a symbiotic star, with
prominent continuum with TiO bands in the red part of the
spectrum. The emission line spectrum includes the HI Balmer series,
HeI~4471, 5875, 6678,7 065, HeII 4686 as strong as \hb\ and faint
5411, faint [OIII]~5007, FeII from multiplet \#42 at 4923, 5018, and
5169~\AA, [FeVII] 4942 5721, [CaV]~6086 or [FeVII]~6087, OI~8446, and
the Raman feature at 6825~\AA. The \ha/\hb\ ratio is 36.

{\it IPHASJ193436.06+163128.9} (Fig.~\ref{F-specsymbio2},
bottom). Superimposed on a red giant continuum with prominent TiO
bands, the emission line spectrum includes HI and HeI lines, HeII
4686, hints of FeII of multiplet \#42, faint OI~8446, and [OIII]~5007
emission about one third the strength of \hb\, and only 2.5 times
larger than [OIII]4363.  The low [OIII]~5007/4363 flux ratio indicates
strong collisional quenching of the [OIII]5007 line at the high
densities typical of S-type symbiotic systems.

{\it IPHASJ193501.31+135427.5} (Fig.~\ref{F-specsymbio3}, top).
Included in the catalogue of \ha\ emission stars of \cite{kw99}, our
spectrum unveils its nature as a symbiotic star. The conclusion is
based on the evidence of shallow absorption bands of an early M giant
similar to IPHASJ184446.08+060703.5, accompanied by an emission line
spectrum of moderately high excitation. This includes lines of the HI
Balmer series, several HeI and FeII lines, weak OI~8446 and HeII~4686,
and forbidden lines of [NeIII]~3869 and [OIII]~5007,4959,4363.  Note
the presence of several strong diffuse interstellar bands (DIBs) in
the spectrum. The [OIII]5007/4363 flux ratios is $\sim$5.


\begin{figure}[!ht]
\centering
\includegraphics{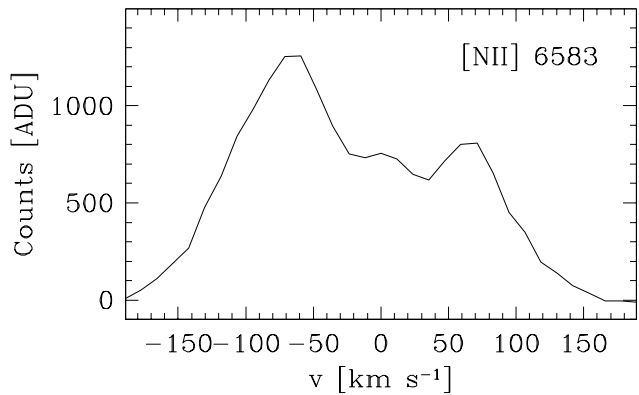}
\caption{The \nii\ line profile of IPHASJ194607.52+223112.3.}
\label{F-specsymbioDhighres}
\end{figure}

{\it IPHASJ194607.52+223112.3} (Fig.~\ref{F-specsymbio3}, bottom).
According to its 2MASS colours, this is the only new D-type symbiotic
star discovered so far by IPHAS (but see other candidates presented in
the next sections).  The object shows an optical spectrum dominated by
nebular lines of low ([OI], [OII], [NII], [SII]) to high ionization
species ([ArIII], [ArV], [OIII], [SIII], HeII, [FeVI], [FeVII],
[CaV]). The \ha/\hb\ ratio is 43.  The continuum is weak and slowly
rising, with no evidence of the absorption bands of a red
giant. However, the presence of the Raman scattered emission at
6825~\AA\ proves its nature as a genuine symbiotic star. The situation
is similar to that of other D-type symbiotic systems, like He 2-104,
where the red giant only shows up in the K band in the near-IR 
(\cite{sg08}).

A spectrum around \ha\ obtained at a higher resolution at the NOT
reveals that each line of the [NII] doublet at 6548 and 6583~\AA\ is
split into at least two components, whose peaks are separated by
130~\kms\ (Fig.~\ref{F-specsymbioDhighres}). A 5~min \nii\ image
obtained at the NOT under a 0$''$.6 seeing does not resolve the nebula
spatially.  In an optically thin case, such a double-peaked profile
cannot be produced by an isotropic mass distribution. In fact,
regardless of the geometrical thickness of the spherical shell, a
substantial fraction of gas emission would be shifted at low or zero
line-of-sight velocity due to projection effects. This would result in
a top flattened line profile. The same conclusion holds for moderately
elliptical shapes. Such a double-peaked profile might instead be
generated by a toroidal, bipolar, or highly collimated mass
distribution inclined at a significant angle from the plane of the
sky.  This is confirmed by simulations covering a range of geometries
and velocity laws, and taking into account the thermal broadening of
the \nii\ line.  However, the [NII]6583/5755 (undereddened) flux
ratios is as low as six, indicating that this expanding region is
dense, with N$_e$ larger than $10^5$~cm$^{-3}$. At such densities, the
optically thin hypothesis might not apply, and simulations of the line
profile would require a more complex modelling that is beyond the
scope of this paper. Note also that the higher resolution spectrum
does not include the \nii~5755 line, and therefore we do not know if
all velocity components in the line are produced by gas in a high
density regime.  We conclude that the object presents an unresolved
dense outflow which is likely to be aspherical, expanding at a speed
larger than $65$~\kms.

\begin{figure}[!hb]
\centering
\includegraphics[height=9.0cm,angle=270]{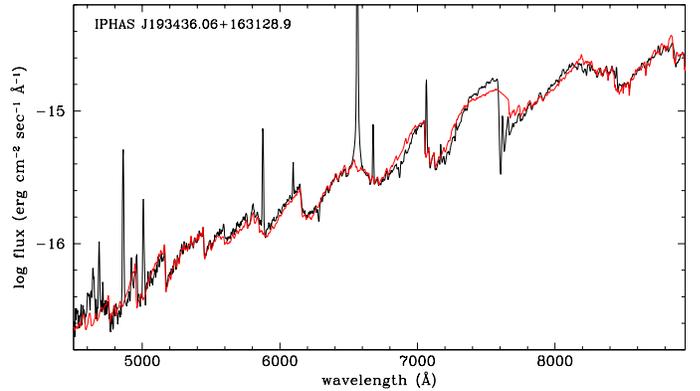}
\caption{Fit (in red) to the observed spectrum (in black) of IPHAS
  J193436.06+163128.9, combining an M5.9~III star, a nebular continuum
  and a $E(B-V)$=1.08 ($R_V$=3.1) reddening.}
\label{fit}
\end{figure}

\subsection{Modelling the red giants and the reddening to derive
the distances}

\begin{figure*}[!ht]
\centering
\includegraphics[width=16cm]{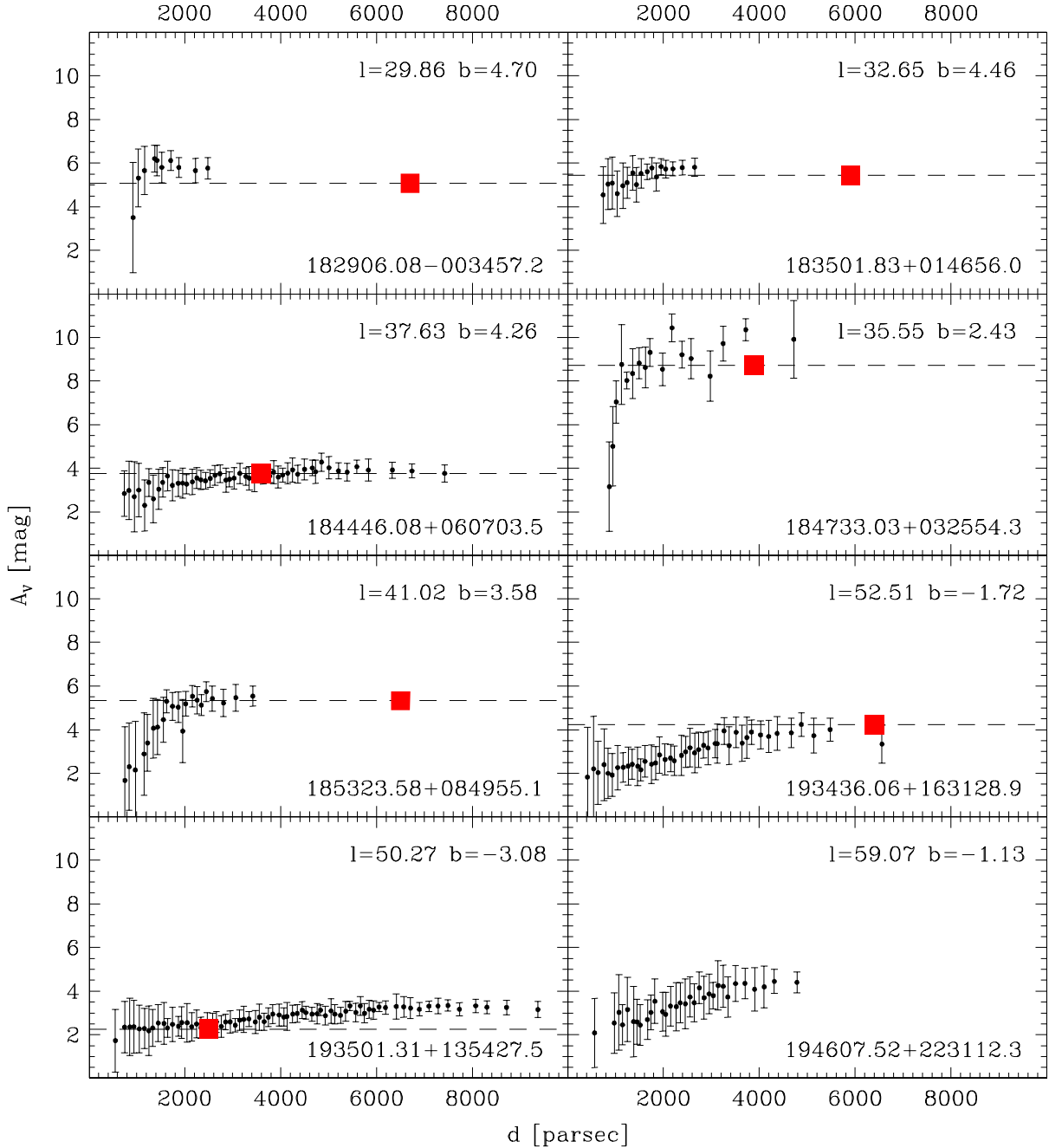}
\caption{Interstellar extinction vs. distance for the sightlines
  toward the new symbiotic stars.  Galactic coordinates are listed in
  the upper-right side of each box. The filled squares and horizontal
  dashed lines indicate distances and reddening determined from the
  fit of the spectra (not available for the object at the bottom-right).}
\label{F-extdist}
\end{figure*}

In order to derive the distances to the new symbiotic stars using the
observed 2MASS K-band magnitudes, the spectral classification and
reddening of their cool giants is needed. They were derived
by fitting the observed spectra. We used the \cite{fl04}
spectral library of red giants and nebular continua from 
\cite{of06}. The R$_V$=3.1 extinction law by Fitzpatrick (1999) was
adopted, and the same amount of reddening was assumed to affect both
the cool giant and the nebular region.  The achieved fits are good and
allow spectral types to be constrained to within 0.3 subtypes and the
$E(B-V)$ reddening to within 0.1 mag. An example of such a fit is
shown in Fig.~5.

The amount of extinction suffered in the V and K bands by the new symbiotic
stars was computed from direct integration over the whole band transmission
profiles (and not simply at their effective wavelengths), following the
procedures outlined in \cite{fm03}. This ensures a higher consistency of the
results. The distance was then worked out assuming that emission in the K
band is entirely due to the red giant, with negligible contribution from
either free-free emission from the nebular component or thermal emission from
circumstellar dust. Based on the infrared colors of symbiotic stars, Whitelock 
and Munari (1992) found that, while it is possible that a few dust-free
symbiotic M stars may be similar to the bright M giants of the solar 
neighborhood, it is clear that as a group the symbiotic stars resemble
the fainter M giants of the Bulge. For this reason, we adopted the
calibration into absolute magnitudes of M spectral types as given by
\cite{fw87}: we used the median values in their Table~3a, scaled to a solar
Galactocentric distance of 8.0~kpc.

The spectral type of the giant, the reddening, the extinction in V and K
bands, the adopted absolute magnitude of the giant and the derived distance
for the new symbiotic stars are summarised in Table~1. This information is
not given for the new D-type symbiotic star IPHASJ194607.52+223112.3 whose
cool giant is not visible in our spectra.

These reddening and distance values are compared with
extinction-distance relationships along the line of sight toward each
object. Indeed, the IPHAS photometry allows the determination
of extinction-distance curves using a large number of main-sequence
field stars. The technique has been presented and discussed by
\cite{s09}. Fig.~\ref{F-extdist} shows the curves for the newly
discovered symbiotic systems, derived from sky areas of $10\arcmin
\times 10\arcmin$ around each line of sight, except for
IPHASJ182906.08-003457.2 and IPHASJ184733.03+032554.3 where the area
considered is twice as large to allow a sufficient number of stars to
be included.  In these two cases, the usual seeing constraint for
IPHAS quality photometry (1$''$.7) was also relaxed to 2$''$.

The distance and extinction derived by fitting the spectra match 
within the errors the growth of the interstellar extinction along
each line of sight. In all but one instance
(IPHASJ193501.31+135427.5), the new symbiotic stars have a reddening
value matching the plateau of the extinction curves produced when the
line of sight breaks out of the thin dust layer in the Galactic
plane. The overall picture of consistency raises our confidence in the
parameters of each system listed in Tab.~\ref{T-symbio}.

\subsection{Two other candidates}

In two other sources, namely IPHASJ190441.53--005957.2 and
IPHASJ190832.31+051226.6 (see Tabs.~\ref{T-yso1} and \ref{T-yso2}), a
red giant continuum and an emission line spectrum were detected. Their
spectra are fit by an M6.8~III star and $E(B-V)$=0.87 reddening, and
by M4-6~III and $E(B-V)$$\sim$1.6, respectively. However, the emission
lines observed are all of low ionization, with only hints of
\oiii\ emission in IPHASJ190832.31+051226.6. 

While completing this article, we noticed that
IPHASJ190832.31+051226.6 brightened 1.8~mag in $r$ from 2004 to 2008
while its \ri\ colour decreased almost one magnitude. This might
indicate the onset of an outburst in the system. We are collecting new
data to further investigate the nature of this interesting object.


\subsection{Objects borderline between symbiotic stars and planetary nebulae}

\begin{figure*}[!ht]
\centering
\includegraphics[width=17.cm]{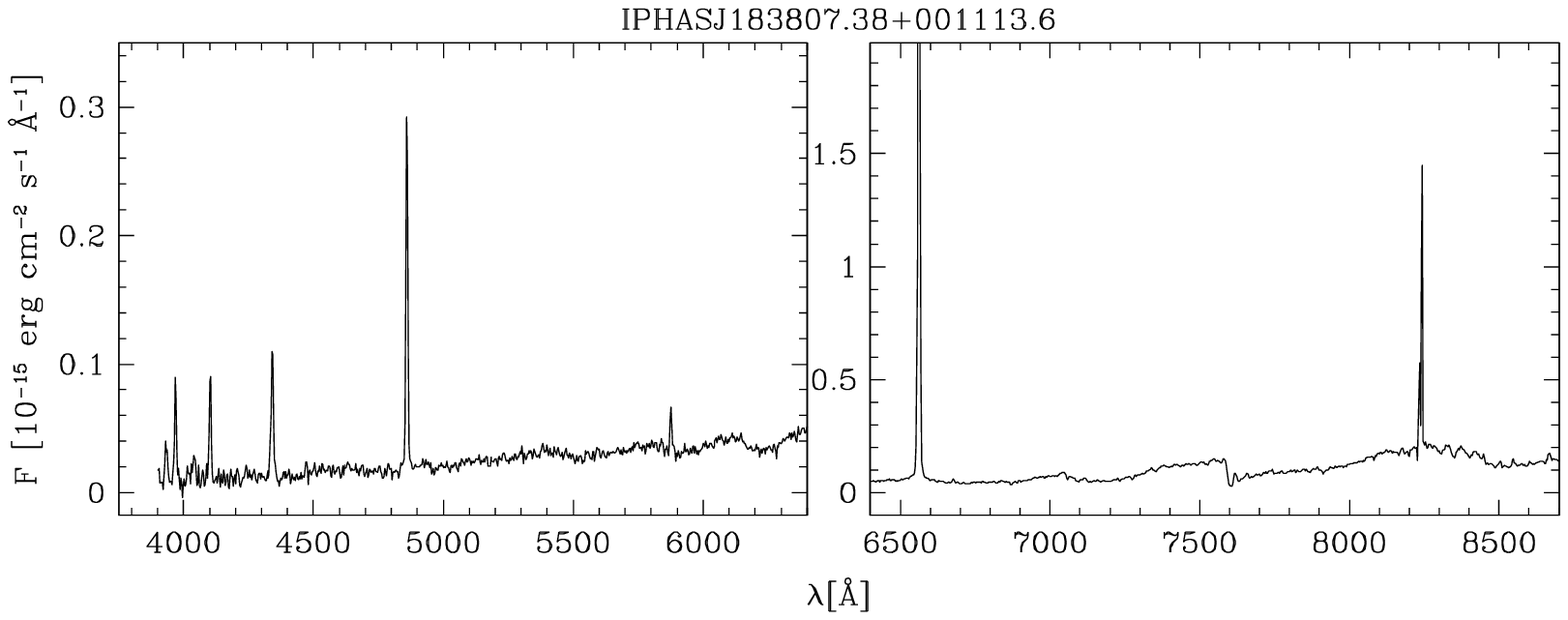}\\[5pt]
\includegraphics[width=17.cm]{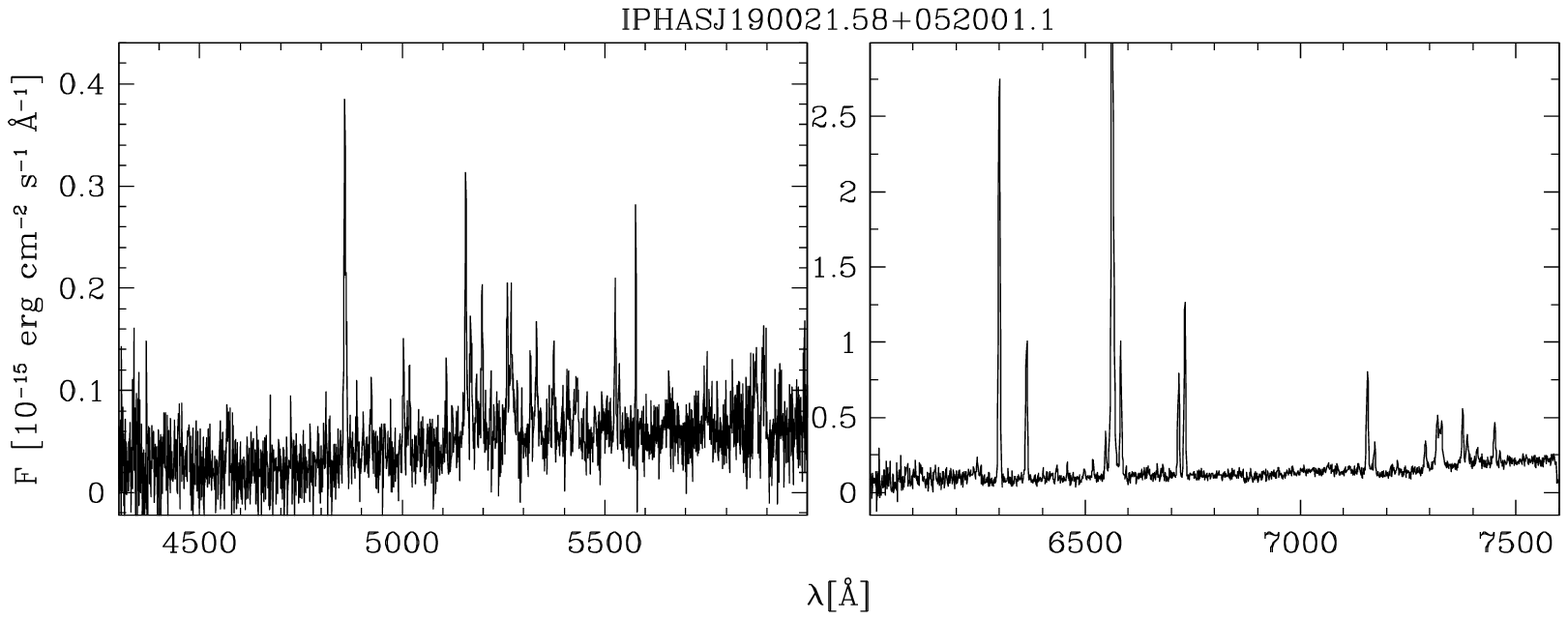}
\caption{Spectra of the CTT star with strong \ha\ emission
  IPHASJ183807.38+001113.6 (top), and of the Be/YSO source with evidence of 
  a compact HII region IPHASJ190021.58+052001.1 (bottom).}
\label{F-yso1}
\end{figure*}

Four objects in our list, namely IPHASJ012544.66+613611.7,
IPHASJ053440.77+254238.2, IPHASJ194907.23+211742.0 and
IPHASJ200514.59+322125.1 (see Tabs.~\ref{T-yso1} and \ref{T-yso2}),
also have a nebular spectrum, as well as 2MASS colours typical of
D-type symbiotic stars. The spectrum of IPHASJ194907.23+211742.0 is
indeed very similar to that of the genuine symbiotic Mira He 2-104
(cf. the atlas of Munari \& Zwitter 2002).  However, the absence of
both the red giant bands and the Raman scattered emission features in
their optical spectra does not allow, at present, a classification as
genuine symbiotic stars. In fact, they are discussed in detail by
Viironen et al. (2009a) and classified as young planetary nebulae.
They form part of a larger sample of objects, including very well
studied nebulae like M~2-9 or Mz~3, whose nature as young planetary
nebulae or symbiotic stars is controversial. The problem and its
complexity have been reviewed by \cite{sg09}.

\section{Other objects}

We now turn to the remaining 48 objects for which we have spectra,
whose properties and classification are also summarised in
Tables~\ref{T-yso1} and \ref{T-yso2}.  Five of them are confirmed as
emission line objects, but are otherwise not yet observed with
sufficient signal to allow a probable object type to be assigned to
them -- these are listed as 'not classified'.  Of the remaining 43
objects, 3 are clearly Wolf-Rayet (WR) stars, and 3 more may be relatively
high-luminosity evolved massive stars.
We also have an emission line object that is
undoubtedly high excitation and helium-rich, may be
associated with an extended nebula visible in the IPHAS images, but
evades classification as a PN.  We say a few more words on these
relatively rare objects below.  A more commonplace finding is
represented by IPHASJ184222.67+025807.0 -- here, it is the only star
we label as dMe.  We base this on its relatively weak H$\alpha$
emission (its equivalent width, at 13~\AA , is the lowest in this
sample), absence of any further indicators of accretion, and lack of
apparent association with a star-forming region.

By far the largest distinctive sub-groupings are of (i) lightly-reddened 
classical T Tauri stars, labelled CTT in Tab.~\ref{T-yso2}, of which there 
are 17, and (ii) more highly reddened, intrinsically brighter emission line 
objects which, in the main, may be either more massive young
stars of HAeBe type or classical Be stars (18 objects, variously
labelled YSO, or Be/YSO).  We look at these more closely first.

\subsection{Young stellar objects and classical Be stars}

Properties common across this group are the presence in their spectra of 
CaII, OI 8446 and permitted FeII line emission -- nearly all objects in
both the CTT and Be/YSO groups present one or more of these, in addition
to strong Balmer line emission.

To earn the designation ``CTT'' in Table~\ref{T-yso2}, the observed
spectrum should show evidence of a late type K/M photosphere, along
with relatively narrow H$\alpha$ emission.  For nearly all the objects
in this group, we give no measurement of H$\alpha$ FWHM for the reason
that the line emission appears unresolved.  For the cases that the
FWHM is quoted, it is modest, between 180 and 350 km~s$^{-1}$, quite
typical of T Tauri stars.  In all cases the H$\alpha$ emission
equivalent width is above the threshold for the rough estimate of
photospheric spectral type at which it is reasonable to infer ongoing
accretion (see e.g. figure 4 in Valdivieso et al. 2009, and Barrado
y Navascues and Mart\'in 2003).

As would be expected, some objects in the CTT group show signs of
continuum veiling in their optical spectra.  In one case
(IPHASJ230342.17+611850.4) the veiling obliterates the underlying
photosphere, but we classify it as CTT on account of the similarity of
its optical spectrum to that of DG Tau (Hessman \& Guenther 1997), and
tentative evidence of LiI~6707 in absorption.  In two further objects
the continuum is well-enough exposed to provide definite detections of
LiI 6707 in absorption. CaII emission is very common in this group, as
previously noticed by \cite{vd08} in the CTTs of Cyg OB2. The NIR
colours of these objects conform with expectation in that they scatter
along the often-drawn T~Tauri locus, and range from exhibiting no
discernable NIR excess to, in one instance, presenting with a very
marked one (see Fig.~\ref{F-cc} and Sect.~5).  Possibly the most
extreme example in this group, deserving of individual mention is
IPHASJ183807.38+001113.6 (Fig.~\ref{F-yso1}, top): it stands out on
account of its extraordinarily high contrast H$\alpha$ (at an
equivalent width of $\sim$500~\AA ), and its M5 photospheric type.

A unifying property of the objects that have been labelled as YSO or
Be/YSO is the presence of a significantly reddened optical continuum,
against which only interstellar absorption is seen (NaI D lines, and a
number of diffuse interstellar bands).  Of course it is not ruled out
that photospheric absorption might become more apparent at appreciably
higher S/N ratio than is typical of the data presently available -
indeed there are 4 objects in which we catch sight of HeI in
absorption.  For 10 out of this set of 18 objects we are able to
resolve the H$\alpha$ profile FWHM and find it ranging from $\sim$200
km~s$^{-1}$ up to $\sim$500 km~s$^{-1}$.  At the high end of this
range, it seems more likely that these objects are main sequence or more
evolved Be objects.  A further consideration that pushes in this
direction is the presence of diffuse interstellar bands at high
equivalent width, since they signal long sightlines through the
diffuse ISM, which in turn implies correspondingly high intrinsic
source brightness.  Conversely, it is possible to be more confident of
a YSO designation where a clear NIR excess is present (see Sect.~5).
But without further data the existing designations have to be viewed
as best guesses.

IPHASJ190021.58+052001.1 (Fig.~\ref{F-yso1}, bottom) is almost in a
class of its own in that, uniquely, its spectrum presents an array of
forbidden lines indicating a compact HII region. There is also a
detectable continuum and breadth to the H$\alpha$ emission line,
likely to be stellar in origin. This star is presumably responsible
for the ionization of the circumstellar environment -- hence its
inclusion under the Be/YSO heading.

\begin{figure*}[!ht]
\centering
\includegraphics[width=14.9cm]{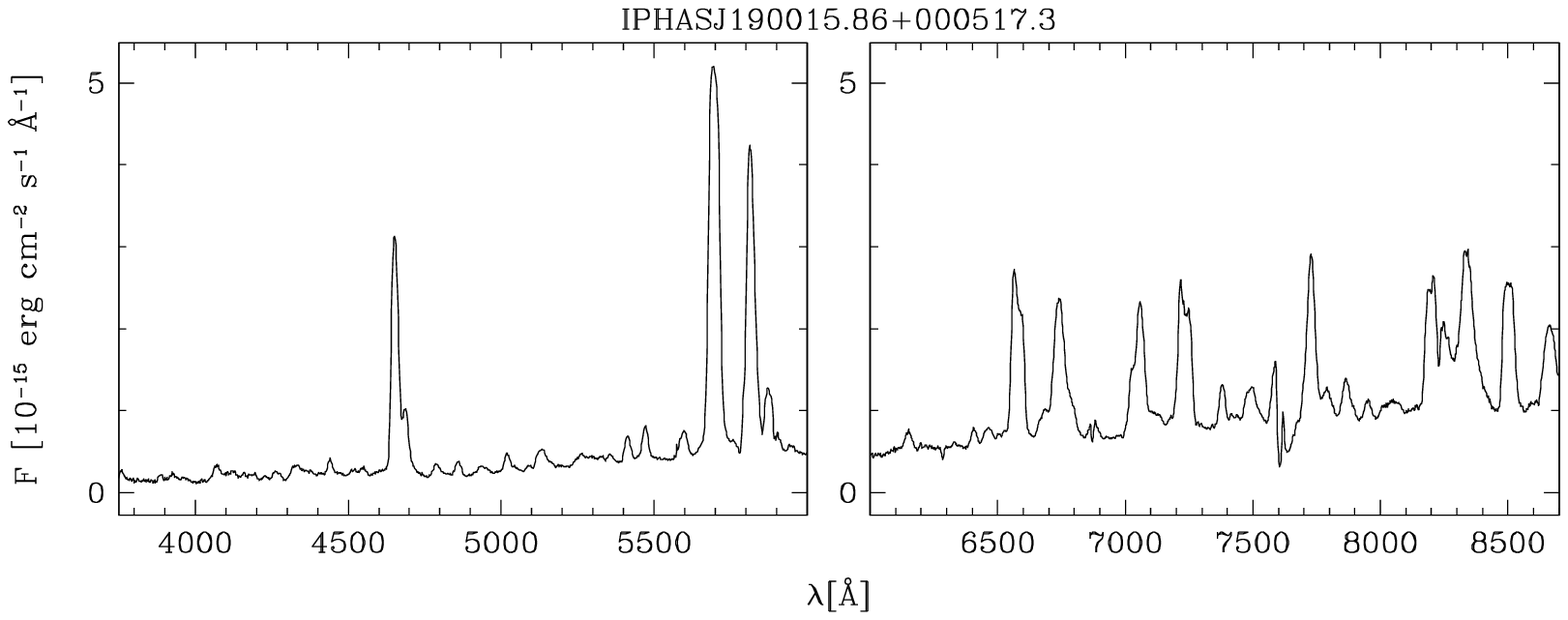}\\[1pt]
\includegraphics[width=14.9cm]{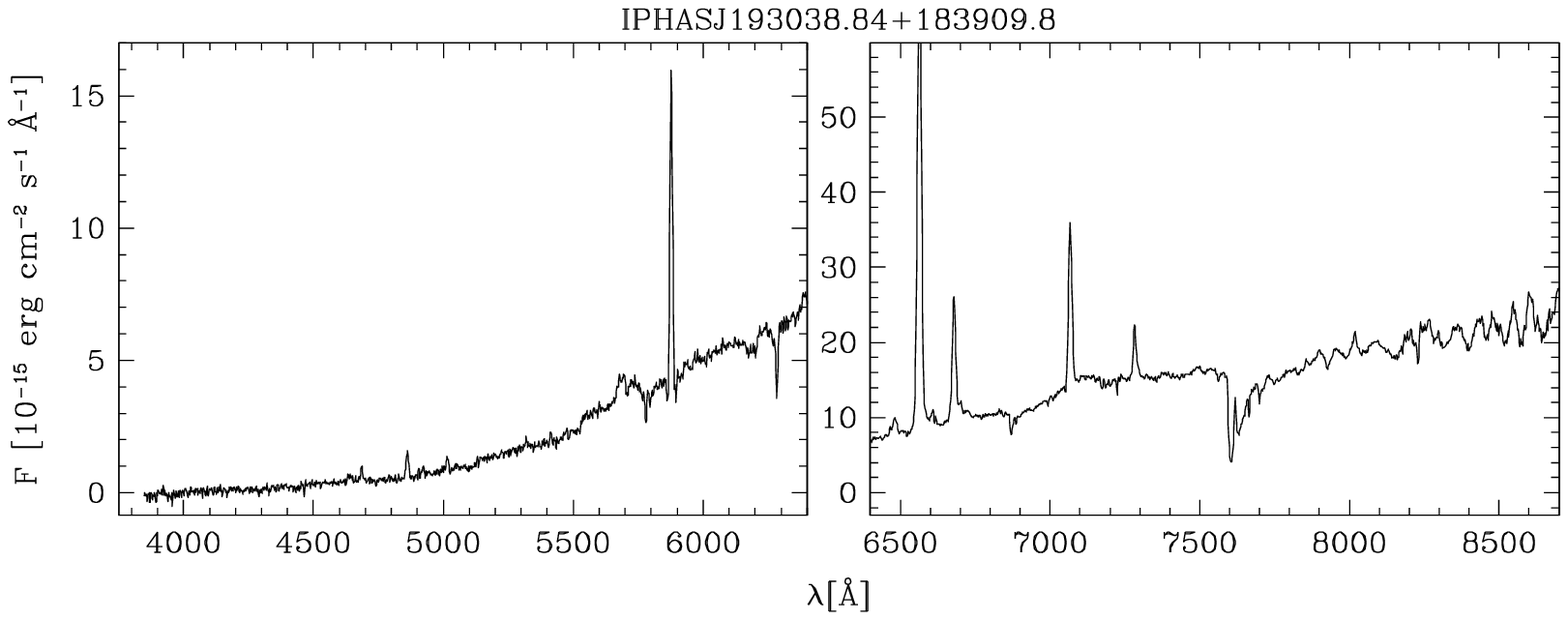}\\[1pt]
\includegraphics[width=14.9cm]{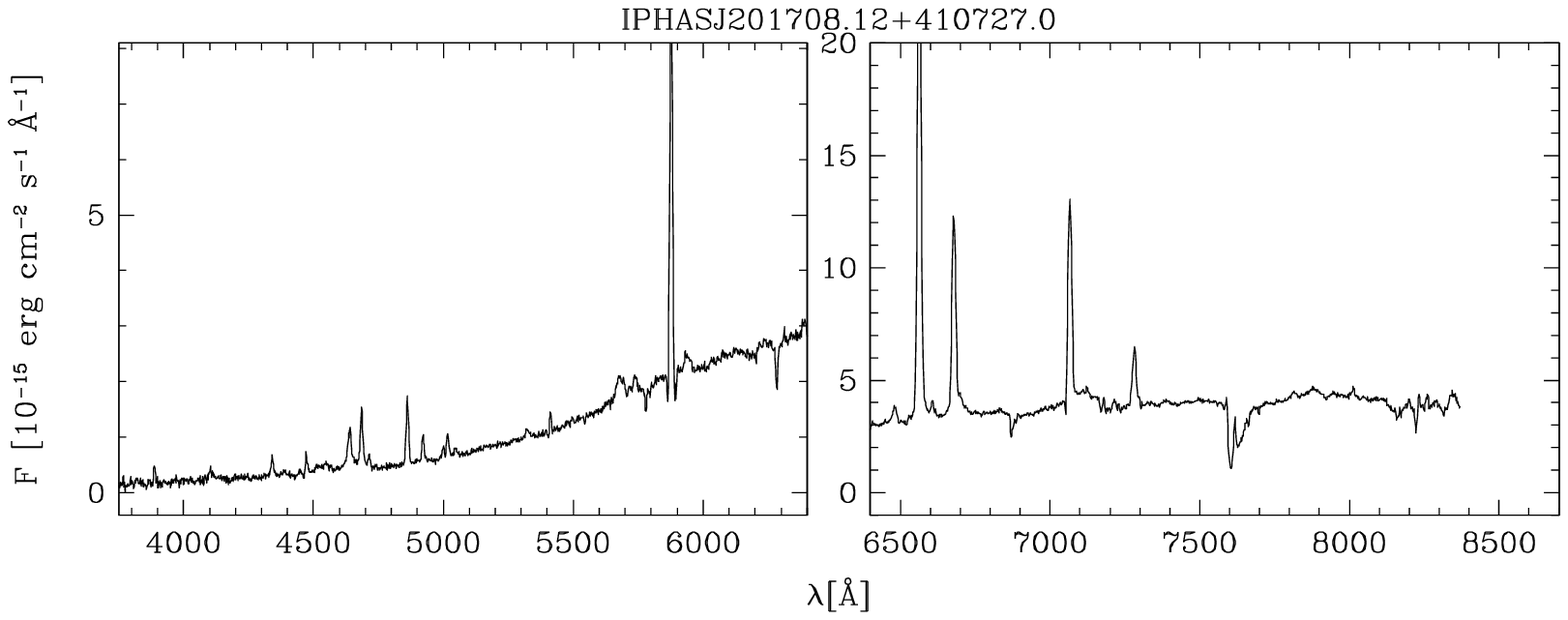}\\[1pt]
\includegraphics[width=14.9cm]{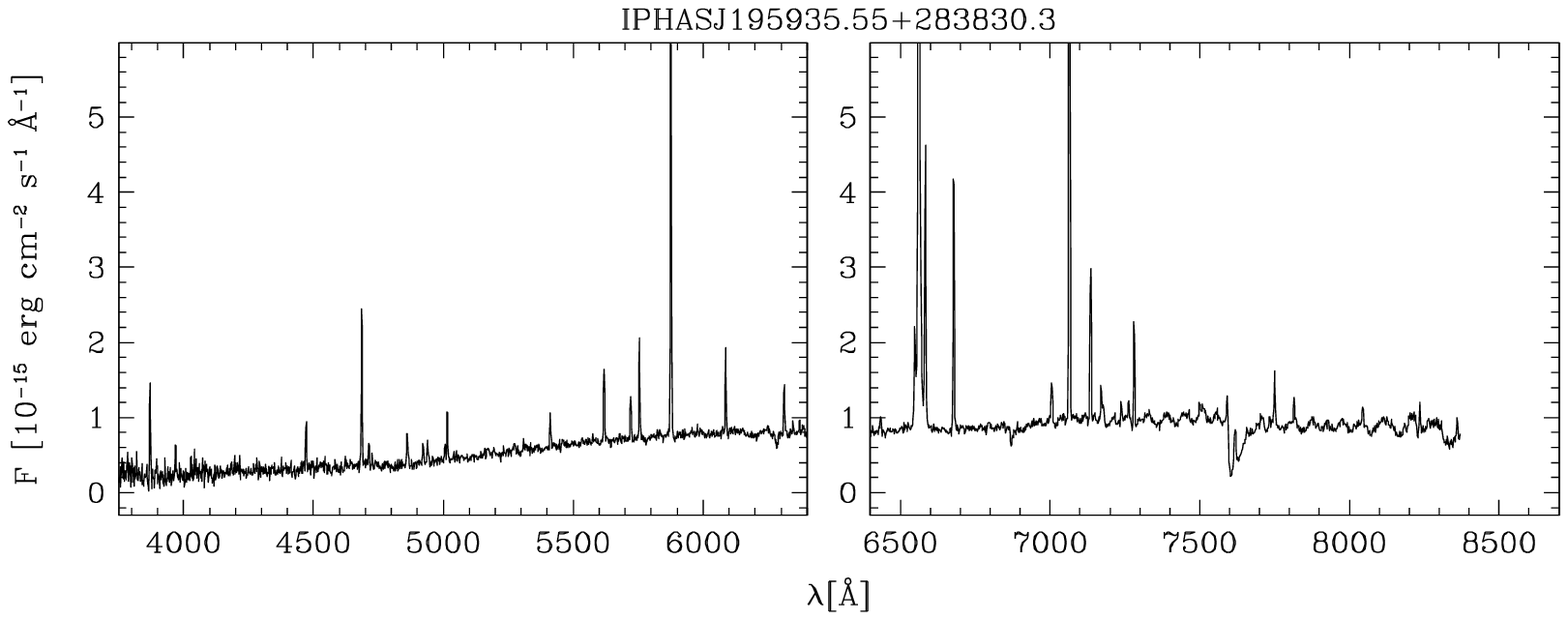}
\caption{Spectra of the Wolf-Rayet stars IPHASJ190015.86+000517.3
  (top), IPHASJ193038.84+183909.8 (second row) and
  IPHASJ201708.12+410727.0 (third row). The spectrum of the
  helium-rich source IPHASJ195935.55+283830.3 (bottom), also possibly related
  to a WR star, is shown in the bottom panel.}
\label{F-yso2}
\end{figure*}

\subsection{IPHASJ190015.86+000517.3, IPHASJ193038.84+183909.8 and 
IPHASJ201708.12+410727.0: three new Wolf-Rayet stars}

IPHASJ190015.86+000517.3 (Fig.~\ref{F-yso2}, top), is a newly discovered
Wolf-Rayet star with a spectrum that can be typed as WC8, based on
comparison with the spectrophotometric library due to Torres \& Massey
(1987). This can be considered as a sort of serendipitous discovery,
as \ha\ is absent and the modest \rha\ excess displayed by the object
is due to the CIV~6560 broad emission.  The Galactic latitude of this
object, $b = -1.92$, takes it further out of the Galactic mid-plane
than is usual for massive WR stars.

In contrast, IPHASJ193038.84+183909.8 and IPHASJ201708.12+410727.0
(Fig.~\ref{F-yso2}, second and third row, respectively) are new WR
stars of WN type.  \cite{g09} have linked the latter object, already
listed as an emission line object by Kohoutek \& Wehmeyer (1999), to a
ring nebula detected at mid-infrared wavelengths.  They have analysed
its optical spectrum and assign a spectral type WN8-9h.  As the
pattern of emission lines present in the optical spectra of both these
stars is very nearly identical, the spectral type of
IPHASJ193038.84+183909.8 must also be WN8-9. \cite{g09} 
deduce the reddening to IPHASJ201708.12+410727.0 (or WR 138a) to be
$A_V = 7.4$.  By comparing the dereddened SED of this object to
progressive dereddening of IPHASJ193038.84+183909.8, we estimate that
the visual extinction of this second WN star is somewhat higher at
$\sim9$ magnitudes.

Note that the WC star we have found lies in a position in the NIR
  diagram (see Fig.~\ref{F-cc}) entirely consistent with the WC locus
  identified by Homeier et al. (2003), who put together data on a
  large sample of WR stars.  The two WN stars also fall in the region
  typical of already-known WN stars.

\subsection{IPHASJ195935.55+283830.3: a helium-rich source surrounded by 
a large nebula}

We show the spectrum of IPHASJ195935.55+283830.3 at the bottom of
Figure~\ref{F-yso2}.  Ahead of in-depth analysis it is difficult to
know how to type this object convincingly.  The emission line
spectrum is quite rich and highly-excited, with all but the \ha\ line
spectrally-unresolved (a Gaussian fit to \ha\ yields a deconvolved
FWHM of $\sim$350 km~s$^{-1}$).  Both HeI and HeII lines are seen in
emission, and it is striking that the HeII 4686 dwarfs \hb.  There is
undoubtedly helium enrichment here.  Nitrogen appears also to be
enhanced, considering that the [OIII] 4959,5007 lines are scarcely
present, while [NII]6548,6583 is strong in this generally
high-excitation spectrum.  On dereddening the SED, assuming $A_V
\simeq 6$, a very blue O-type continuum is recovered, but no obvious
stellar features are seen with the present resolution and S/N.  The
NIR excess of this source is also remarkable (see Tab.~\ref{T-yso1}).

\begin{figure}[!h]
\centering
\includegraphics[width=9.1cm]{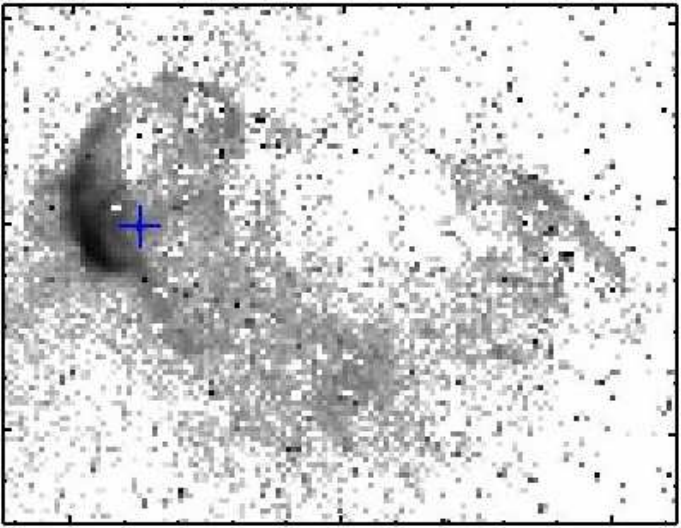}
\caption{The continuum-subtracted \ha\ IPHAS image of the nebulosity
  around IPHASJ195935.55+283830.3. The position of the star is indicated by the
  cross. 5$''$$\times$5$''$ pixel binning has been applied to
  highlight the faintest structures in the nebula. The field of view is 
 650$''$$\times$500$''$. North is up, East is left.}
\label{F-neb195935}
\end{figure}

This object is associated with a faint, asymmetrical ionized nebula
with a major axis that is nearly 10 arcmin long
(Fig.~\ref{F-neb195935}). In particular, the star is located near the
bright, bow-shaped eastern edge of the nebulosity.

We note that the spectrum of IPHASJ195935.55+283830.3 is very similar
to that of NaSt1 (\cite{cs99}), except that in the latter \ha\ is
relatively narrow (FWHM$\sim$50~\kms). These authors suggest that such
a spectrum is produced in dense photoionized ejecta consisting of
fully CNO-processed gas, which prevents the observation of a
Wolf-Rayet central star of early WN or WC type. This is the reason why
IPHASJ195935.55+283830.3 is marked as ``WR nebula?''  in
Tab.~\ref{T-yso2}. However, alternative hypotheses should be explored,
like the possibility that the source is a young compact binary
emerging from a PN phase. A deeper analysis, including additional data
that we are acquiring for both the spatially-unresolved emission-line
source and the surrounding large nebula, will be presented in a
forthcoming paper.

\subsection{IPHASJ205544.33+463313.5, and IPHASJ190229.97-022757.0, two B[e] 
stars}

The spectra of IPHASJ205544.33+463313.5 and IPHASJ190229.97-022757.0
(Fig.~\ref{F-yso3}, top and middle, respectively)
are notable for their low excitation permitted and forbidden lines,
including for the latter, fluorescent [NiII].  Both exhibit very
pronounced NIR excesses also (see Table~\ref{T-yso1} and Sect.~5)
and enormously high-contrast H$\alpha$ emission.  This combination of
properties is what has been dubbed the 'B[e] phenomenon' (Lamers et al.
1998).  The presentation of [NiII] fluorescent emission, identified
via the ratios among the 6666, 7378 and 7412 line, is not very common
-- other locations noted for it include the LBV P~Cygni and the Orion
Nebula (see Lucy 1995 for a discussion of the fluorescent process).
It appears to signal strongly illuminated circumstellar environments.
For IPHASJ190229.97-022757.0, the rough 6666:7378:7412 flux ratio
pattern is 0.17:1:0.44 -- to be compared with the pattern for P Cygni
(Barlow et al. 1994) that is 0.13:1:0.27.

\subsection{IPHASJ185005.71-004041.2: low-mass or massive star?}

\begin{figure*}[!ht]
\centering
\includegraphics[width=17.cm]{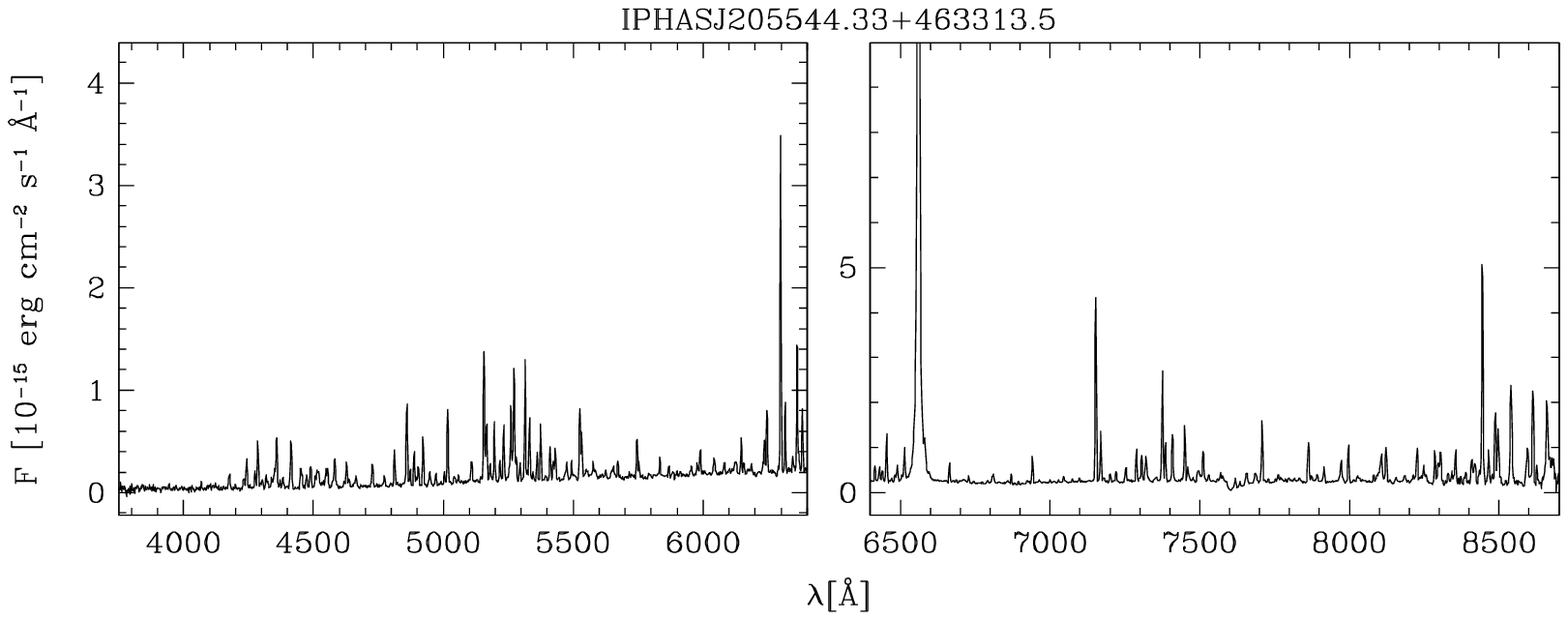}\\[5pt]
\includegraphics[width=17.cm]{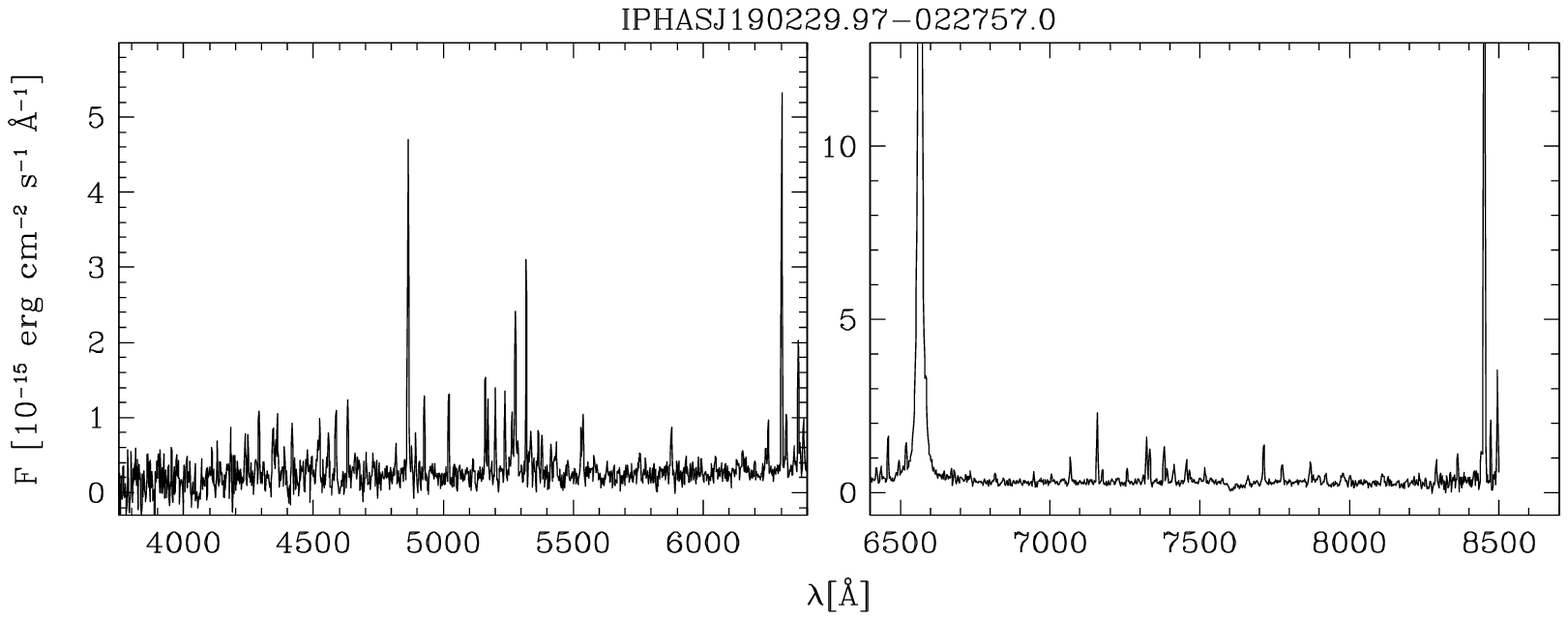}\\[5pt]
\includegraphics[width=17.cm]{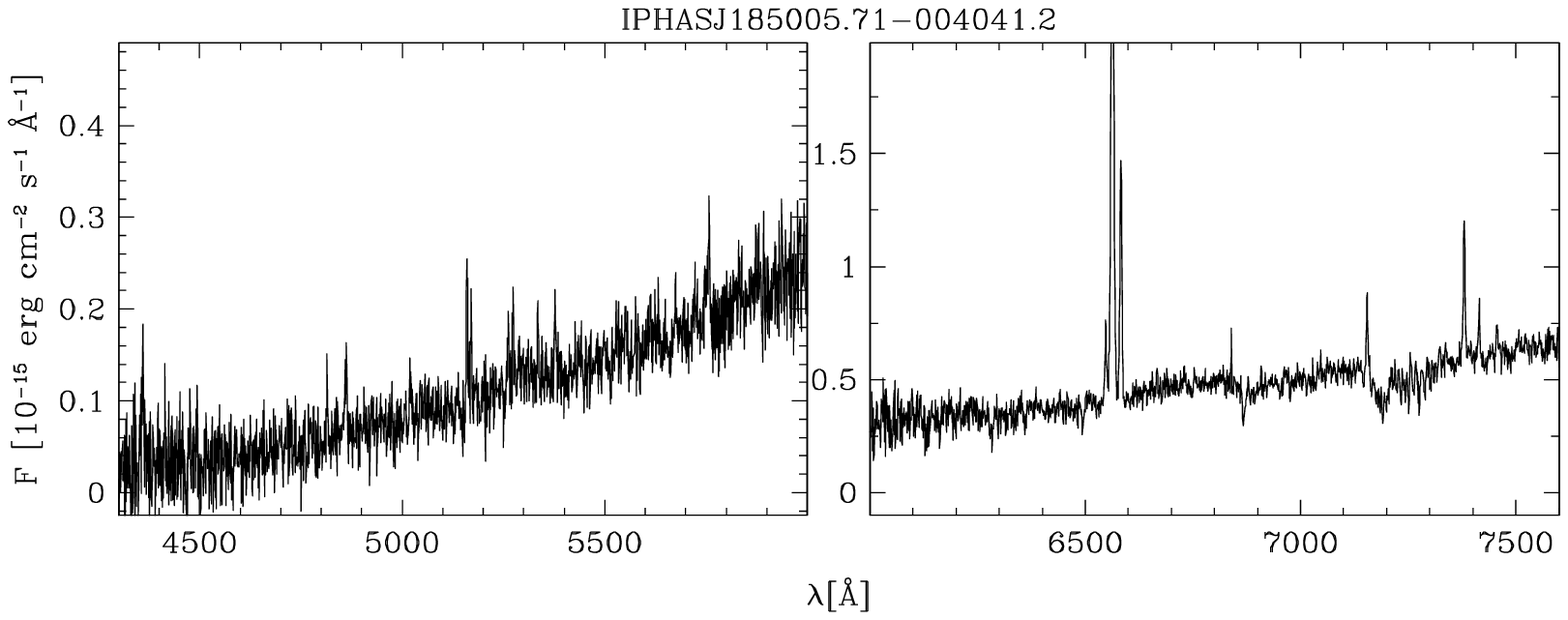}
\caption{Spectra of the B[e] stars IPHASJ205544.33+463313.5 (top) and
  IPHASJ190229.97-022757.0 (middle), and of the source of uncertain nature  
  but with similar [NiII] fluorescent emission 
  IPHASJ185005.71-004041.2 (bottom).}
\label{F-yso3}
\end{figure*}

IPHASJ185005.71-004041.2 (Fig.~\ref{F-yso3}, bottom) has a much more
subdued emission line spectrum than the B[e] stars discussed in the
previous section.  However, it shares with IPHASJ190229.97-022757.0
the presence of [NiII] fluorescent emission.  The 6666:7378:7412 flux
ratio pattern in this case is even closer to the P~Cygni prototype
(0.12:1:0.29). Quite marked [NII] emission is seen in this object
also, but no HeI is apparent.  

Understanding the nature of this source with the present data is
difficult.  The presence of [NiII] fluorescent emission would suggest
that it might also be a massive evolved star, perhaps a LBV. But the
[NiII] flux ratio also points to low reddening ($A_V\le4$~ mag) for
what is a highly reddened mid-plane sightline ($\ell = 32.18$, $b =
-0.01$), which would make it relatively close and thus intrinsically
faint. On the other hand, the detection of reasonably strong DIB
absorption and the 2MASS colours of the object suggest instead a
significant reddening, i.e. a relatively distant luminous
object. Given the uncertainty on its absolute magnitude, until a
higher quality spectrum is available no classification is proposed in
Tab.~\ref{T-yso2} for this star.

\section{Discussion and perspectives}
\label{S-diagram}

\begin{figure*}[!ht]
\centering
\includegraphics[width=14cm]{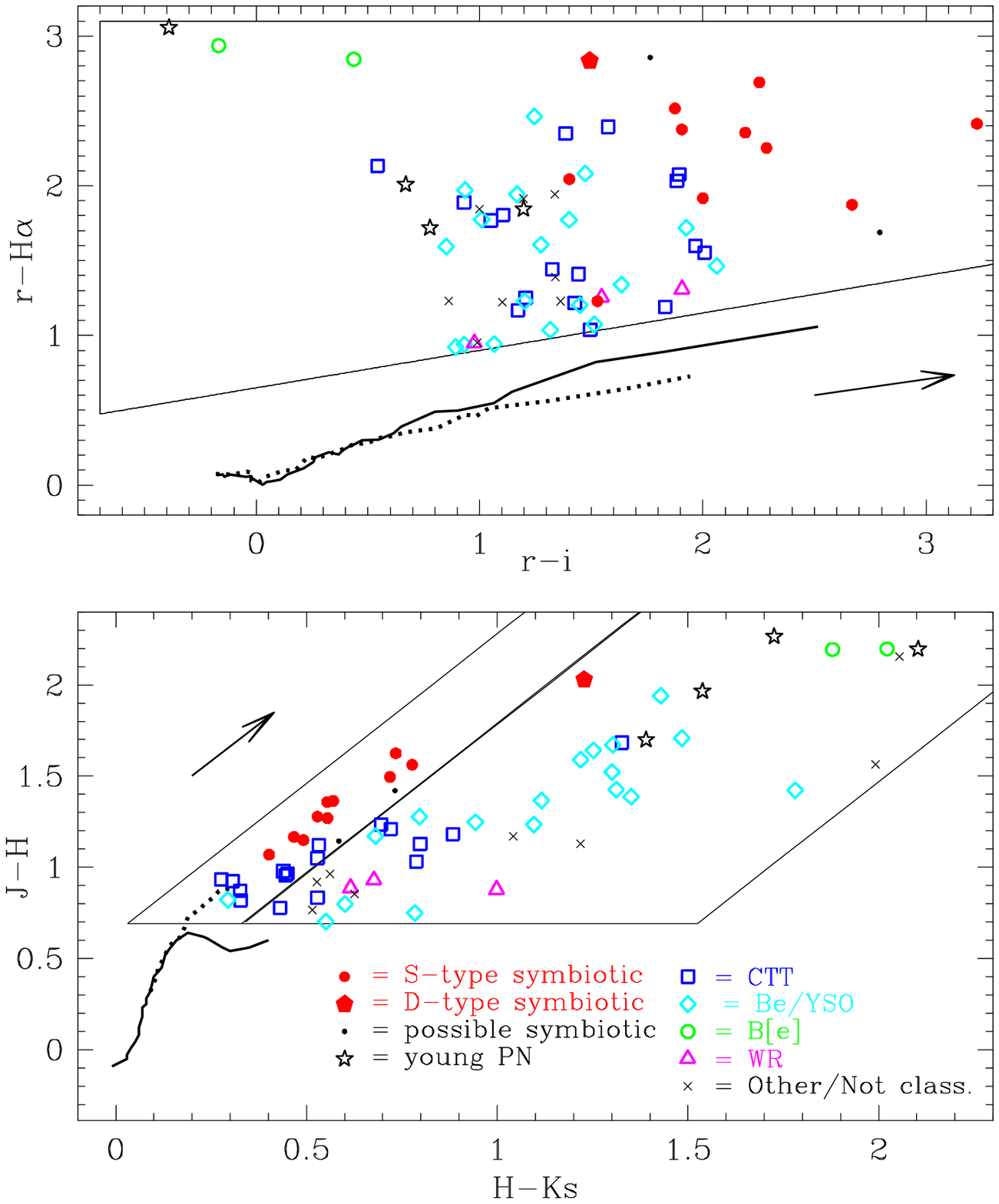}
\caption{IPHAS (top) and 2MASS (bottom) colour-colour diagrams for 
the objects observed spectroscopically. The three new IPHAS symbiotic
stars from paper I are also included.  The locus of unreddened
main-sequence and RGB stars are indicated by the solid and dotted
lines, respectively.
The arrow indicates the reddening vector for M0 giants: its length
corresponds to 3~mag extinction in V.  The symbiotic stars selection
boxes as defined in paper I are also indicated. In the IPHAS diagram,
the boxes are common for both types. In the 2MASS diagram, the box for
the D-types is to the right side of the contiguous box for the
S-types.}
\label{F-cc}
\end{figure*}

The IPHAS and 2MASS colour-colour diagrams for the sample of sources
observed spectroscopically, including the three new symbiotic
stars presented in paper I, are shown in Fig.~\ref{F-cc}.  The new
symbiotic stars are generally displaced toward the upper-right side of
the IPHAS diagram compared to the other \ha\ emitters  
that we have observed. In the 2MASS diagram, the S-type
symbiotic stars form an isolated clump corresponding to reddened cool
giants. T Tauri and other young stars are much more dispersed in both
diagrams, and are generally found at smaller \rha\ colours than
symbiotic systems.  

Comparison with Figs.~1 and 2 of paper I shows that the new symbiotic
stars from IPHAS generally have redder colours -- both in the optical
and in the near-IR -- than the sample of objects of this class which
were previously known. They are also optically fainter: their $r$ 
magnitudes are between 14.0 and 18.5, compared to the sample of known
objects considered in paper I, which were mostly filling the magnitude
interval between r=9.5 and r=14.5 mag. The luminosity difference is
smaller if the K band is considered, which encourages the conclusion 
that the redder colours and fainter optical magnitudes are mainly due to
the greater interstellar reddening, encountered along many of the lines of
sight in the Galactic plane explored by IPHAS. Thus, as expected
IPHAS seems to mainly discover reddened objects, which previously
escaped detection because of their relatively faint optical magnitude.
The IPHAS symbiotic stars discovered so far are
concentrated between R.A.=18.5~h (close to the southern limit
where the IPHAS coverage starts) and R.A.=20.5~h, i.e. at Galactic
longitudes between 30 and 81 degrees. These are directions towards the
inner Galaxy: considering the distances in Tab.~\ref{T-symbio}, we
conclude that at least half of the new symbiotic stars are located on
the side of the bulge closer to the Sun.

Together with the three objects of paper I, so far we have found
eleven new symbiotic systems in the Galactic plane. 
Considering that before our survey only eleven systems were known in
the IPHAS area (Belczy\'nski et al. 2000), we have doubled the number
of symbiotic stars known in this region of the sky.  The
result is positive, but we are still orders of magnitude below the
large number of objects predicted to exist in the Galaxy (see
Sect. 1). In order to obtain a new, reliable estimate of the total
Galactic population of symbiotic stars, the following steps have to be
followed:\\[-18pt]
\begin{itemize}
\item
Build a complete sample of candidates.\\ The possible symbiotic stars
in paper I were selected from the list of IPHAS \ha\ emitters by
\cite{w08}, which in turn was built when the survey's photometric
catalogue was only partially completed.  We now need to produce a
complete, magnitude limited sample of candidates, scanning the whole
IPHAS photometric catalogue in a similar way as done by Viironen et
al. (2009b) for compact PNe.  A similar search is foreseen in the
opposite hemisphere when the IPHAS southern extension, VPHAS+, will be
carried out at the 2.6mVST in Chile. The latter will also cover
part of the Galactic bulge, where a large fraction of symbiotic stars
are located.
\item
Estimate accurately our success rate for the detection of symbiotic
stars from the photometric selection.\\ So far, including the results
in paper I, our success rate for S-type symbiotic stars is close to
50\%\ (10 out of 22 candidates observed spectroscopically). As paper I
lists 337 candidates, we might expect to find another 150 new S-type
systems if we continue the spectroscopic follow up. However, the newly
discovered systems have redder \ri\ and \jh\ colours than the majority
of the candidates (cf. Fig. 6 of paper I), most of which are also
quite faint.  Thus if the remaining objects prove to be symbiotic
stars, they would be faint red giants with little reddening,
i.e. faraway objects in clear lines of sight, a combination that is
unlikely in the Galactic plane. Indeed, all the objects located in the
lowest part of the selection box for S-types observed
spectroscopically, turned out to be young stars. We conclude that our
global success rate will be significantly lower than obtained so
far. More spectra are needed in order to determine this.

We have found only one new D-type system. Spectroscopy confirms the
idea discussed in paper I that the vast majority of candidates in the
adopted D-type selection boxes are young or massive stars. As D-type
systems represent 15--20\%\ of the known sample of Galactic symbiotic
stars, the effect of missing them is not dramatic and can be taken 
into account via a correction factor.
\item
Transform the estimate of the total number of objects in the IPHAS
area to the total number in the Galaxy.\\ This implies introducing the
results of our search (completeness and success rate as a
function of magnitude and direction) into a model that includes the
spatial distribution of symbiotic stars in the Galaxy and the
growth of interstellar extinction in every direction.
\end{itemize}

\begin{acknowledgements}

We are grateful to many of our collaborators in the IPHAS project, for
continuous discussion about the properties of the variety of objects
that are involved in the analysis of the survey data. RLMC, ERRF, AM,
and KV acknowledge funding from the Spanish AYA2007-66804 grant.  We
are grateful to the Mount Stromlo and Siding Spring Observatory Time
Allocation Committee for enabling the spectroscopic follow-up to be
obtained.

\end{acknowledgements}

\begin{table*}[!h]
\caption{IPHAS magnitudes (from Witham et al. 2008) and 2MASS
  magnitudes of non-symbiotic stars.}
\begin{tabular}{lcccrrr}
\hline\hline
Name (IPHASJ....)  & r      & i   & \ha    &\multicolumn{1}{c}{J} &\multicolumn{1}{c}{H} &\multicolumn{1}{c}{K}  \\
                   &  \multicolumn{1}{c}{[mag]} &     &        &        &      &   \\
\hline&         \\[-7pt]                                                                                                
000432.31+580854.0 & 17.36 & 16.02 & 15.41 & 13.63 & 12.86 & 12.35 \\
012544.66+613611.7 & 18.87 & 18.20 & 16.86 & 15.48 & 13.22 & 11.49 \\
031704.34+601500.0 & 16.65 & 15.33 & 15.21 & 13.33 & 12.35 & 11.91 \\
032039.49+562358.2 & 14.68 & 13.40 & 13.07 & 10.95 &  9.57 &  8.22 \\
035823.95+522312.6 & 18.96 & 17.77 & 17.05 & 15.74 & 14.83 & 14.30 \\
045625.15+434931.8 & 18.13 & 17.13 & 16.29 & 15.25 & 14.08 & 13.04 \\ 
053018.12+313558.9 & 16.95 & 16.02 & 16.01 & 14.32 & 13.50 & 13.21 \\
053113.16+382006.6 & 16.53 & 15.68 & 14.94 & 12.54 & 11.74 & 11.14 \\
053440.77+254238.2 & 17.48 & 16.28 & 15.63 & 14.09 & 12.12 & 10.58 \\
055254.08+171424.7 & 16.47 & 15.54 & 14.58 & 13.86 & 13.03 & 12.50 \\
060515.15+204036.7 & 17.82 & 16.48 & 16.43 & 14.77 & 13.21 & 11.22 \\
060821.91+295255.7 & 17.80 & 16.35 & 16.39 & 14.41 & 13.37 & 12.84 \\
183748.03-001617.2 & 16.94 & 15.51 & 15.72 & 13.25 & 12.32 & 12.04 \\
183807.38+001113.6 & 18.47 & 17.08 & 16.12 & 14.20 & 13.24 & 12.79 \\
183814.63-012213.8 & 17.30 & 15.30 & 15.75 & 12.16 & 10.98 & 10.10 \\
184222.67+025807.0 & 16.03 & 15.04 & 15.08 & 13.32 & 12.36 & 11.80 \\
184431.46-001652.4 & 17.21 & 15.76 & 16.00 & 12.54 & 10.87 &  9.56 \\
184635.85+005521.4 & 14.86 & 13.69 & 13.70 & 11.73 & 10.04 &  8.72 \\
185005.71--004041.2& 16.95 & 15.59 & 15.72 & 13.00 & 11.87 & 10.66 \\
185349.55+052353.7 & 18.48 & 17.08 & 16.70 & 14.34 & 13.06 & 12.26 \\
185424.82+041905.0 & 18.29 & 16.46 & 17.10 & 13.13 & 11.92 & 11.20 \\
185448.29+005033.5 & 18.32 & 16.43 & 16.25 & 13.57 & 12.45 & 11.92 \\
190015.86+000517.3 & 14.94 & 13.97 & 13.99 & 12.18 & 11.30 & 10.30 \\
190021.58+052001.1 & 17.90 & 16.73 & 15.95 & 13.73 & 12.09 & 10.83 \\
190229.97--022757.0& 16.13 & 16.30 & 13.19 & 13.42 & 11.22 &  9.34 \\
190441.53--005957.2& 15.88 & 13.08 & 14.19 &  8.89 &  7.75 &  7.16 \\
190832.31+051226.6 & 17.79 & 16.02 & 14.93 & 10.68 &  9.26 &  8.53 \\
190857.31+053620.6 & 16.89 & 15.82 & 15.95 & 12.83 & 11.47 & 10.35 \\
191017.43+065258.1 & 15.82 & 13.89 & 14.10 & 10.42 &  8.71 &  7.22 \\
192033.79+231040.3 & 16.89 & 15.39 & 15.85 & 13.43 & 12.65 & 12.22 \\
192249.80+142236.3 & 16.96 & 15.32 & 15.62 & 12.61 & 11.19 &  9.88 \\
192400.05+230253.1 & 15.79 & 14.68 & 14.56 & 12.82 & 11.97 & 11.34 \\
192515.05+224720.3 & 16.40 & 15.29 & 14.60 & 13.38 & 12.57 & 12.24 \\
192841.27+174819.9 & 16.70 & 15.49 & 15.47 & 13.67 & 12.43 & 11.48 \\
193038.84+183909.8 & 13.26 & 11.35 & 11.95 &  8.21 &  7.28 &  6.60 \\
193232.88+151711.5 & 14.16 & 13.27 & 13.24 & 11.73 & 10.98 & 10.20 \\
194609.56+225423.2 & 15.20 & 13.69 & 14.12 & 11.50 & 10.79 & 10.24 \\
194907.23+211742.0 & 16.77 & 15.99 & 15.05 & 13.58 & 11.88 & 10.49 \\
195935.55+283830.3 & 15.81 & 14.95 & 14.58 & 12.48 & 10.32 &  8.27 \\
200514.59+322125.1 & 17.24 & 17.63 & 14.18 & 14.96 & 12.76 & 10.66 \\
201057.51+343732.4 & 15.68 & 14.36 & 14.64 & 12.08 & 10.84 &  9.75 \\
201152.98+281517.0 & 17.26 & 15.31 & 16.00 & 12.31 & 10.71 &  9.50 \\
201708.12+410727.0 & 14.29 & 12.74 & 13.04 & 10.15 &  9.27 &  8.65 \\
201918.94+381448.5 & 18.61 & 17.68 & 16.64 & 16.04 & 14.87 & 14.19 \\
202834.25+355417.4 & 15.89 & 14.42 & 13.81 & 11.50 & 10.08 &  8.29 \\
202947.93+355926.5 & 14.75 & 13.74 & 12.98 & 10.96 &  9.44 &  8.14 \\
203413.39+410157.9 & 19.00 & 17.75 & 16.54 & 13.23 & 11.29 &  9.86 \\
203545.40+400332.6 & 19.46 & 17.50 & 17.86 & 14.93 & 13.96 & 13.51 \\
204134.22+393239.0 & 15.16 & 14.11 & 13.39 & 12.12 & 10.99 & 10.20 \\
205544.33+463313.2 & 15.72 & 15.28 & 12.87 & 12.61 & 10.41 &  8.39 \\
210404.87+535124.4 & 18.21 & 16.63 & 15.82 & 13.43 & 12.40 & 11.61 \\
222628.63+612049.0 & 18.68 & 18.13 & 16.54 & 14.62 & 13.75 & 13.42 \\
230342.17+611850.4 & 17.55 & 15.66 & 15.52 & 13.04 & 11.81 & 11.12 \\
231334.92+620944.1 & 17.12 & 15.91 & 15.87 & 14.13 & 13.21 & 12.90 \\    
\hline
\end{tabular}
\label{T-yso1}
\end{table*}

\longtabL{3}{
\begin{landscape}
\begin{longtable}{llllrrl}
\caption{\label{T-yso2}Main spectral characteristics of non-symbiotic
  stars and proposed classification.  The spectral range covered by our
  spectra is indicated, for reference for future studies of the
  objects. For line identification, the symbol ``:'' indicates
  marginal or uncertain detection. The \ha\ FWHM, corrected for the
  instrumental profile, is given in all cases where the
  spectral resolution allowed us to resolve the line width.}\\
\hline\hline
Name (IPHASJ....)  & Spectral    &  Continuum and       & Main emission features                & \multicolumn{2}{c}\ha  & Classification \\
                   & range       &  absorption features &                                       & EW    &   FWHM         &       \\
                   &       [nm]  &                      &                                       & [\AA] &  [\kms]        &       \\
\hline\\[-7pt]                                                                                                
\endfirsthead
\caption{continued.}\\
\hline\hline
Name (IPHASJ....)  & Spectral    &  Continuum and       & Main emission features                & \multicolumn{2}{c}\ha  & Classification \\
                   & range       &  absorption features &                                       & EW    &   FWHM         &       \\
                   &       [nm]  &                      &                                       & [\AA] &  [\kms]        &       \\
\hline\\[-7pt]                                                                                                
\endhead
\hline
\endfoot
000432.31+580854.0 & 370-750 & featureless             & HI,HeI,FeII,\cab                          & 195 & 350 & Not classified    \\
012544.66+613611.7 & 370-835 & negligible              & nebular$^\star$,OI8446,\car                & 950 & --  & young PN$^\star$\\
031704.34+601500.0 & 390-780 & TiO                     & HI, FeII,\cab                             & 122 & --  & CTT \\
032039.49+562358.2 & 370-870 & featureless,DIB,NaI D   & HI,FeII,OI7772,8446,\car:                 & 210 & --  & YSO     \\
035823.95+522312.6 & 390-745 & featureless             & HI,FeII,HeI:                              & 200 & 250 & Not classified  \\
045625.15+434931.8 & 600-740 & featureless             &  \ha                                      & 60  & --  & Not classified \\ 
053018.12+313558.9 & 370-780 & featureless,DIB         & HI,HeI                                    & 45  & --  & Be/YSO      \\
053113.16+382006.6 & 535-745 & featureless,DIB+NaI     & \ha                                       & 63  & 500 & Be/YSO  \\
053440.77+254238.2 & 370-870 & featureless,DIB         & nebular$^\star$,OI8446,\cabr               & 200 & 300 & young PN$^\star$\\
055254.08+171424.7 & 350-900 & faint TiO               & HI,Balmer jump,HI,HeI,FeII,\cabr          & 240 & --  & CTT? \\
060515.15+204036.7 & 600-740 & featureless             & \ha                                       &  62 & 300 & Not classified        \\
060821.91+295255.7 & 384-855 & TiO                     & HI,HeI,FeII,faint OI8446,\cabr            & 110 & 230 & CTT    \\
183748.03-001617.2 & 370-900 & faint TiO               & HI,faint HeI                              & 40  & 280 & CTT   \\
183807.38+001113.6 & 390-900 & TiO                     & HI,HeI,\cabr                              & 500 & --  & CTT \\
183814.63-012213.8 & 370-820 & TiO                     & \ha,\hb,[OI],\cab                         & 175 & 250 & CTT \\
184222.67+025807.0 & 370-870 & faint TiO,strong NaI    & \ha                                       & 13  & --  & dMe \\
184431.46-001652.4 & 390-900 & featureless,HeI6678,NaI,DIB & HI,FeII,OI8446,\car                   &  78 & 400 & Be/YSO     \\
184635.85+005521.4 & 370-900 & Li I 6707,6495,NaI,DIB  & HI,FeII,faint HeI,OI8446,[SII],[OI],\cabr &  95 & 230 & CTT     \\
185005.71--004041.2& 425-775 & featureless,DIB         & HI,FeII,[FeII],[NII],[NiII]6666,7378,7412 &  65 & 250 & ? (see text) \\
185349.55+052353.7 & 370-900 & featureless             & HI,HeI,FeII,\car                          & 215 & --  & YSO    \\
185424.82+041905.0 & 600-900 & weak TiO                & HI,strong \car                            &  38 & 350 & CTT \\
185448.29+005033.5 & 370-890 & TiO                     & HI,HeI,[OI]                               & 90  & --  & CTT         \\
190015.86+000517.3 & 370-900 & rising                  & dominated by strong broad C lines         & --  & --  & WC8      \\
190021.58+052001.1 & 430-773 & featureless             & HI,FeII,[FeII],[OI],[OII],[NII],[SII]     & 180 & 250 & Be/YSO   \\
190229.97--022757.0& 370-850 & faint continuum         & HI,HeI,FeII,[FeII],[OI],[OII],OI8446      & 1500& 250 & B[e]       \\
190441.53--005957.2& 370-900 & deep TiO                & HI,FeII,faint HeI                         &  70 & 250 & symbiotic?   \\
190832.31+051226.6 & 370-850 & TiO                     & HI,HeI,OI8446,[OIII]:                     & 250 & 280 & symbiotic?   \\
190857.31+053620.6 & 370-900 & featureless,DIB,NaI D   & HI,[NII],OI8446                           & 90  & 350 & Be/YSO               \\
191017.43+065258.1 & 370-900 & featureless,HeI5876,DIB,NaI D & HI,HeI6678(PCyg),[OI],OI8446,\car   & 150 & 350 & YSO               \\
192033.79+231040.3 & 430-775 & TiO                     & HI                                        &  35 & 100 & CTT?      \\
192249.80+142236.3 & 430-775 & featureless,DIB         & \ha,[OI]                                  &  100& 330 & Be/YSO               \\
192400.05+230253.1 & 370-900 & featureless             & HI,faint \cab                             & 42  & 300 & Not classified     \\
192515.05+224720.3 & 390-900 & TiO                     & HI,HeI,\cab                               & 80  & --  & CTT \\
192841.27+174819.9 & 370-900 & featureless             & HI,HeI,FeII,complex \ha,OI8446,strong \car&  63 & 200 & YSO               \\
193038.84+183909.8 & 370-900 & featureless,DIB         & strong HeI, HeII                          &  120 & 700& WN8-9                 \\
193232.88+151711.5 & 370-900 & HI,HeI,strong DIB,NaI   & HI,FeII,OI8446,\car                       &   52 & 400& Be/YSO                \\
194609.56+225423.2 & 380-840 & featureless,DIB,NaI D   & \ha,\hb                                   &  55  & 250&  Be/YSO               \\
194907.23+211742.0 & 370-900 & faint                   & nebular$^\star$                            & 250 & --  & young PN$^\star$\\
195935.55+283830.3 & 360-835 & featureless,DIB         & strong HeI and HeII,HI,[NII],[NeIII],[OIII],& 80 & 350  & WR nebula? \\
                   &         &                         & [FeVII],[CaVII],[SIII],[ArIII],[ArIV],[ArV] &    &      &               \\
200514.59+322125.1 & 370-856 & negligible              & nebular$^\star$,OI8446                     & 4000& --  & young PN$^\star$\\
201057.51+343732.4 & 370-870 & featureless,strong DIB  & HI,FeII,OI7772,8446,\car                  &  77 & --  & YSO?                 \\
201152.98+281517.0 & 370-870 & featureless             & HI, OI 8446, strong \car                  &  85 & 220 & YSO?                 \\
201708.12+410727.0 & 370-870 & featureless, DIB        & HI,broad strong HeI and HeII              &  95 & 700 & WN8-9              \\
201918.94+381448.5 & 370-880 & featureless             & HI,HeI,faint \cab                         & 380 & --  & YSO             \\
202834.25+355417.4 & 372-875 & featureless,HeI5876,Na ID,DIBS& HI,[OI],OI8446,\car                 & 370 & 200 & YSO                \\
202947.93+355926.5 & 370-870 & featureless,weak HeI    & HI,FeII,OI8446,\car:                      & 180 & 350 & YSO$^\dag$\\
203413.39+410157.9 & 510-870 & featureless             & \ha,[OI],OI8446                           & 370 & 300 & YSO       \\
203545.40+400332.6 & 370-900 & TiO                     & HI                                        & 105 & --  & CTT          \\
204134.22+393239.0 & 370-870 & faint TiO,LiI 6707      & HI,HeI,FeII,[SII],strong \cabr            & 160 & 200 & CTT   \\
205544.33+463313.2 & 373-875 & faint                   & rich FeII,[FeII],HI,faint HeI,OI8446,[Ni II],\car  &1500 & 180 & B[e] \\
210404.87+535124.4 & 370-860 & TiO                     & HI,HeI,OI8446                             & 350 & 180 & CTT          \\
222628.63+612049.0 & 370-900 & TiO                     & HI,HeI                                    & 260 & --  & CTT          \\
230342.17+611850.4 & 370-900 & featureless             & HI,FeII,\cab,[SII],strong \car,OI7772,8446& 145 & --  & CTT          \\
231334.92+620944.1 & 390-780 & TiO,NaI D               & HI                                        &  80 & --  & CTT         \\
\end{longtable}
$^\star$ Viironen et al. (2009b)

$^\dag$  Krause et al. (2003)
\end{landscape}
}

\end{document}